%% file: graphm.tex
\documentclass[aps,pre,twocolumn,groupedaddress,showpacs,floatfix]{revtex4}
\usepackage{amsmath}
\usepackage{amssymb}
\usepackage{graphicx}
\usepackage{epic}
\usepackage{eepic}

\begin{document}

\title{An Analytical Approach to Connectivity in Regular Neuronal
Networks}

\author{Luciano da Fontoura Costa and Marconi Soares Barbosa}
\affiliation{Institute of Physics of S\~ao Carlos.
University of S\~ ao Paulo, S\~{a}o Carlos,
SP, PO Box 369, 13560-970,
phone +55 162 73 9858,FAX +55 162 71
3616, Brazil, luciano@if.sc.usp.br}

\date{\today}

\begin{abstract} 
This paper describes how realistic neuromorphic networks can have
their connectivity fully characterized in analytical fashion.  By
assuming that all neurons have the same shape and are regularly
distributed along the two-dimensional orthogonal lattice with
parameter $\Delta$, it is possible to obtain the exact number of
connections and cycles of any length from the autoconvolution function
as well as from the respective spectral density derived from the
adjacency matrix.  It is shown that neuronal shape plays an important
role in defining the spatial distribution of synapses in neuronal
networks. In addition, we observe that neuromorphic networks typically
exhibit an interesting phenomenon where the pattern of connections is
progressively shifted along the spatial domain for increasing
connection lengths.  This is a consequence of the fact that in neurons
the axon reference point usually does not coincide with the cell
centre of mass. Morphological measurements for characterization of the
spatial distribution of connections, including the adjacency matrix
spectral density and the lacunarity of the connections, are suggested
and illustrated. We also show that Hopfield networks with connectivity
defined by different neuronal morphologies, quantified by the proposed
analytical approach, lead to distinct performace for associative
recall, as measured by the overlap index.  The potential of the
proposed approach is illustrated with respect to digital images of
real neuronal cells.

\end{abstract}


\maketitle

\section{introduction}

A particularly meaningful way to understand neurons is as cells
optimized for \emph{selective connections}, i.e. connecting between
themselves in a specific manner so as to achieve proper circuitry and
behavior.  Indeed, the intricate shape of dendritic trees provide the
means for connecting with specific targets while minimizing both the
cell volume and the implied metabolism (e.g. \cite{Purves:1985,
Karbowski:2001}).  While great attention has been placed on the
importance of synaptic strength over the emerging neuronal behavior,
geometrical features such as the shape and spatial distribution of the
involved neurons are closely related to the network connectivity.  In
addition, the topographical organization and connections pervading the
mammals' cortex provide further indication that adjacencies and
spatial relationships are fundamental for information processing by
biological neuronal networks.  The importance of neuronal geometry has
been reflected by the growing number of related works (see, for
instance, \cite{SI_BM:2003}).  However, most of such approaches target
the characterization of neuronal morphology in terms of indirect and
incomplete measures such as area, perimeter and fractal dimension of
the dendritic and axonal arborizations.  Interesting experimental
results regarding the connectivity of neuronal cells growth \emph{in
vitro} have been reported in \cite{Shefi:2002, Segev:2003} and what is
possibly the first direct computational approach to neuronal
connectivity was only recently reported in \cite{Percolation:2003},
involving the experimental estimation of the critical percolation
density as neuronal cells are progressively superposed onto a
two-dimensional domain.  At the same time, the recent advances in
complex network formalism (e.g. \cite{Albert_Barab:2002,
Bollobas:2002, Barabasi_Ravasz:1998, Buckley:1990, Amaral:2000})
provide a wealthy of concepts and tools for addressing connectivity.
Initial applications of such a theory to bridge the gap between
neuronal shape and function were reported in
\cite{Stauffer:2003,Costa_BM:2003}.

As the investigation of the relationship between neuronal shape and
function is underlined by computational approaches involving
numerical methods and simulation, a need arises to develop an
analytical framework for neuromorphic characterization that could lead
to additional insights and theoretical results regarding the
relationship between neuronal shape and function.  The present work
describes an analytical approach capable of fully characterizing the
connectivity of morphologically-realistic neuronal networks composed
by repetitions of the same neuron along space.  Such a kind of network
can be considered as a model of biological neuronal systems
characterized by planarity and morphologic regularity, as is the case
with ganglion cell retinal mosaics \cite{Wassle:1974} and the basal
dendritic arborization of cortical pyramidal cells. The basic idea
underlying the proposed analytical approach is to use the symmetries
induced by the periodical boundary conditions in order to allow the
connecting matrix describing the network to become a circulant matrix.
Important features such as the number of connections and cycles can
then be exactly obtained from the spectrum of this matrix.  Lacunarity
is also considered as a complementary measurement of the connectivity
pattern.  The effect of different neuronal shapes over the dynamics of
the respective neuronal systems (Hopfield) built upon such connections
is then investigated by using the proposed methodology. It is shown
that neuronal shape not only plays an important role in defining the
spatial distribution of synapses in neuronal networks, but also
imposes critical constraints over the respective behavior.

\section{Methodology}

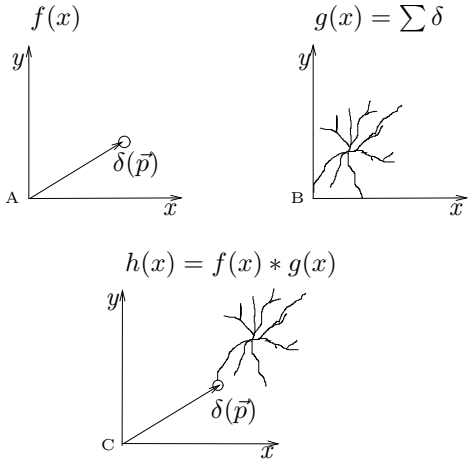
\begin{figure}
\begin{center}
\input{prop.eepic}
  \caption{The convolution of an axon function and a dendritic 
   destribution~\label{fig:prop}}
\end{center}
\end{figure}

The analytical representation of the connectivity of a neuronal
network developed in the current section is based on the convolution
of a function representing the neuronal cell with Dirac deltas.  The
basic adopted construction is illustrated in Figure~\ref{fig:prop},
where the convolution of the neuronal cell $g(x)$ in (b) with the
Dirac delta $f(x)$ in (a) produces as effect a copy of the original
cell at the position of the delta (c).  This can be mathematically
expressed as:

\begin{equation}
  \delta(\vec{a}) \ast g(\vec{x}) = g(\vec{x}-\vec{a})
\end{equation}

Let the neuronal cell be represented in terms of the triple
$\eta=[A,S,D]$ where $A$ is the set of points belonging to its axonal
arborization, $S$ is the set of points corresponding to the respective
soma (neuronal body) and $D$ are the dendritic arborization points.
For simplicity's sake, a finite and discrete neuronal model is
considered prior to its continuous general formulation.  We therefore
assume that the points used to represent the neuron belong to the
square orthogonal lattice $\Omega= \left\{ 1, 2, \ldots, N \right\}
\times \left\{ 1, 2, \ldots, N \right\}$, with parameter $\Delta =1$.
The axon and soma are represented by a single point each.  Such points
could be understood as corresponding to the tip of the axon and the
soma center of mass, respectively.  The dendritic arborization is
represented in terms of the finite set of dendrite points $D={D_1,
D_2, \ldots, D_M}$ and, in order to prevent the formation of loops, it
is henceforth assumed that a dendrite point never coincides with the
axon.  Figure~\ref{fig:neurmod} illustrates such a geometrical
representation for a neuron with 3 dendrite points.  Observe that the
coordinate origin coincides with the axon, which is taken as reference
for the soma and dendrite coordinates.  Observe that the arrows in
this figure refer to the relative positions of the soma and dendrites
and not to the signal transmission by a real neuronal cell, which
occurs in the opposite direction (i.e. from dendrites to axon).


\begin{figure}
\begin{center}
\input{neuron_def.eepic}
\caption{The geometry of a simplified neuronal cells represented in
terms of its axon $A$, soma centroid $S$ and dendrite points
$D_i$.~\label{fig:neurmod}} \end{center}
\end{figure}
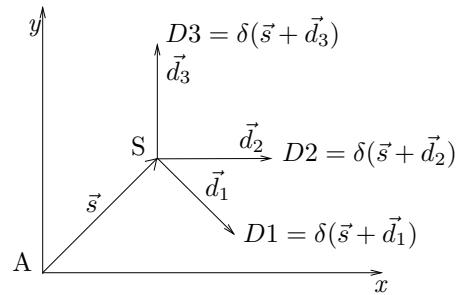

\begin{figure}
\begin{center}
\input{diagram.eepic}
  \caption{The neurons connected through unit-length paths to neuron $i$
  which is placed with its axon $A$ at $\vec{p}$ can be obtained through the
  convolution between the position $\delta(\vec{p})$ of neuron $i$ with the
  dendritic structure function.~\label{fig:constr}}
\end{center}
\end{figure}
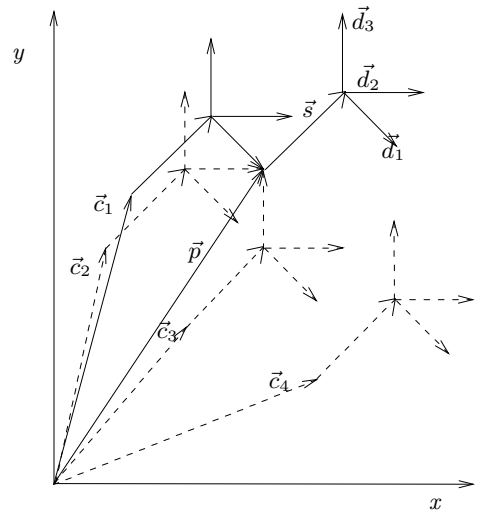

Neuromorphic networks (actually digraphs \cite{Bollobas:2002}) can now
be obtained by placing one such a neuron at all possible nodes of the
orthogonal lattice $\Omega$.

A connection is established whenever an axon is overlaid onto a
dendrite point. The thus obtained connection pattern stands out as a
particularly important feature of the obtained network, as it defines
the possible communications between the cells.  Consequently, it is
important to obtain analytical expressions describing the respective
neuronal connectivity, e.g. by considering the spatial distribution of
paths and cycles of any specific length along the network.

We start by considering the connections with a single specific neuron
$i$ placed with its axon at position $\vec{p}$, illustrated in
Figure~\ref{fig:constr} together with four other neuronal cells at
positions $\vec{c_1}$, $\vec{c_2}$, $\vec{c_3}$ and $\vec{c_4}$.
Given the particular geometry of the basic cell, three connections are
implied with the cells at $\vec{c_1}$, $\vec{c_2}$ and $\vec{c_3}$
whose dendrites coincide with the axon of the cell $i$.  In the case
of cell $\vec{c_1}$, this situation can be mathematically expressed as
$\vec{c_1}=\vec{p}-\vec{d_1}-\vec{s}$.  The fourth cell $\vec{c_4}$
illustrates one of the many neurons which are \emph{not} connected to
cell $i$.  Two directly connected cells are henceforth represented as
two nodes of a graph connected by a unit-length path (a simple arc).

As is clear from such a construction, the set of neurons connected to
$i$ through unit-length paths can be obtained by convolving the
initial point~\footnote{Unless mentioned otherwise we will be using
the simplifying notation for the dirac delta function
$\delta(x-a)=\delta(a)$} $\delta(\vec{p})$ (the tip of the axon) with
the function $g(x,y)$ representing the neuronal cell 

\begin{equation}
 g(x,y)=\delta(-\vec{d}_1-\vec{s}) + \delta(-\vec{d}_2-\vec{s}) +
     \delta(-\vec{d}_3-\vec{s}).
\end{equation}

Observe that the minus signs in this equation are implied by the need
to flip the function representing the cell shape along both axes, so
as to obtain proper propagation of the connections.  For instance, in
Figure~\ref{fig:constr}, the propagation of information proceeds from the
original axon at $\vec{p}$ to the dendrites at $\vec{c_1}$,
$\vec{c_2}$, and $\vec{c_3}$.

More generally given a set of initial neurons with axons represented
as a Dirac's delta distribution $\xi(x,y)$, the density of dendrites
connected to those neurons by unit-path lengths can be obtained from
Equation~\ref{eq:chi}.  Observe that $\xi(x,y)$ may contain Dirac deltas
with intensity larger than one, resulting from sums of coinciding
deltas. The function $\nu(x,y)$ in Equation~\ref{eq:nu} is analogous
to $\xi(x,y)$ but here all Dirac deltas have unit intensity.  The
functions expressing the density of the dendrites connected to the
original neurons through paths of length $k$ is given by
Equation~\ref{eq:chik}. The number of connections with length $k$ is
given by Equation~\ref{eq:tau}. Observe that the use of the Dirac delta
function in such a formulation allows the immediate extension of such
results to continuous spatial domains.

While the analytical characterization of the connectivity of the
considered network model has been allowed by the fact that identical
neuronal shapes are distributed along all points of the orthogonal
lattice, it is interesting to consider extensions of such an approach
to other situations.  An immediate possibility is to consider sparser
configurations, characterized by larger lattice parameters $\Delta$.
Such an extension involves sampling the neuronal cell image at larger
steps.  

\begin{equation}
 \chi(x,y)=g(x,y) \ast \xi(x,y) \label{eq:chi} \\ 
\end{equation}

\begin{equation}\label{eq:nu}
\nu(x,y)=
      \begin{cases}
      \delta(x,y) & if \quad \xi(x,y) \neq 0 \\
      0 & otherwise.
      \end{cases}  
\end{equation}

\begin{eqnarray}
\chi_k(x,y)=\underbrace{ g(x,y) \ast \ldots \ast g(x,y) }_{k  \times}
\xi(x,y) \label{eq:chik} \\  
\tau_k(x,y)=\sum_{j=1}^{k}(\chi_j(x,y)) \label{eq:tau}
\end{eqnarray}

\begin{figure}
\begin{center} \vspace{0.3cm}
 \includegraphics[angle=-90,scale=0.5]{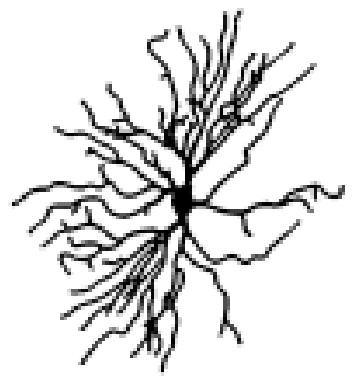} \hspace{1.5cm}
 \includegraphics[angle=-90,scale=0.5]{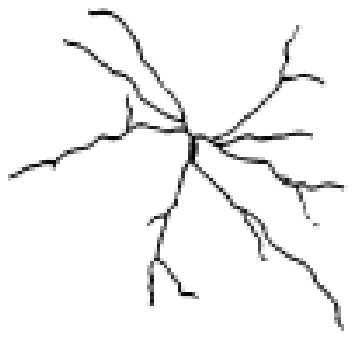} \\
 \vspace{0.2cm} (a) \hspace{2.5cm} (b) \\
 \includegraphics[angle=0,scale=0.4]{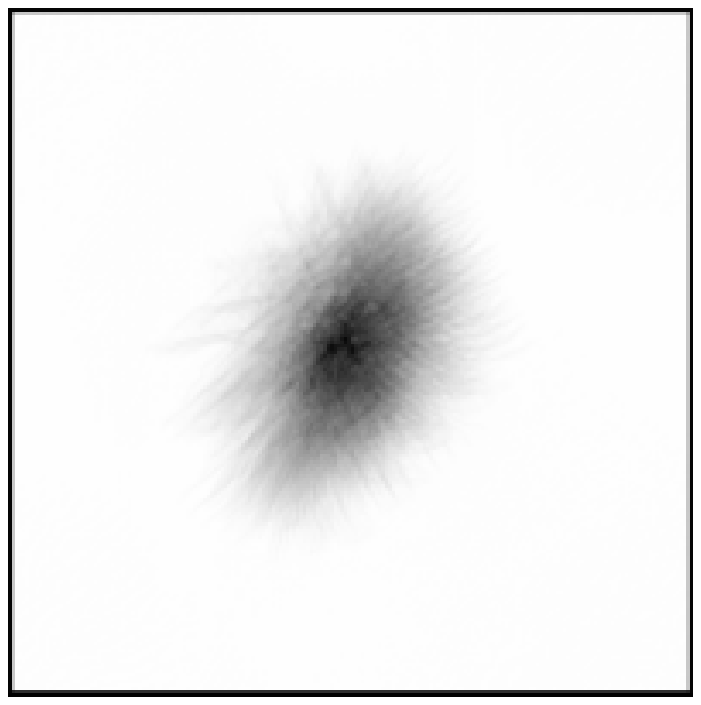} \hspace{0.5cm}
 \includegraphics[angle=0,scale=0.4]{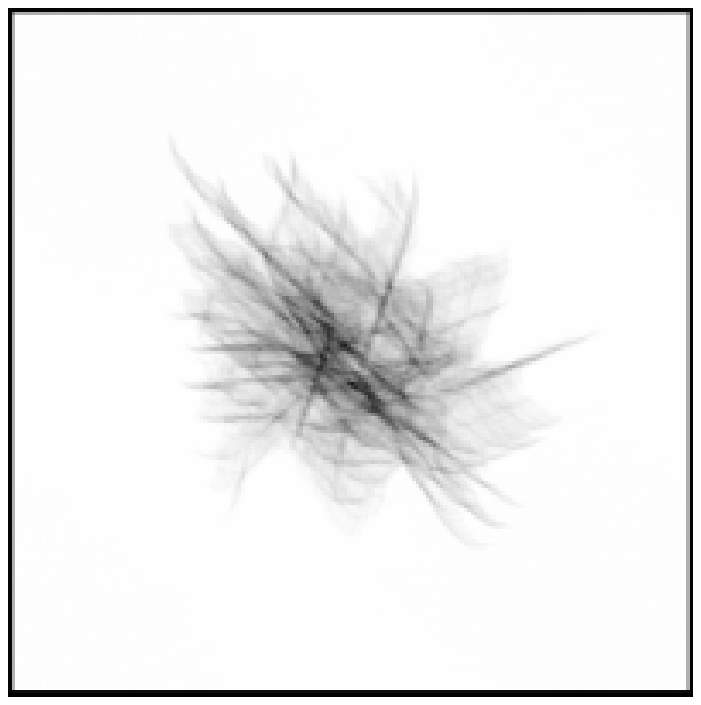} \\
 \vspace{0.2cm} (c) \hspace{2.5cm} (d) \\
 \includegraphics[angle=0,scale=0.4]{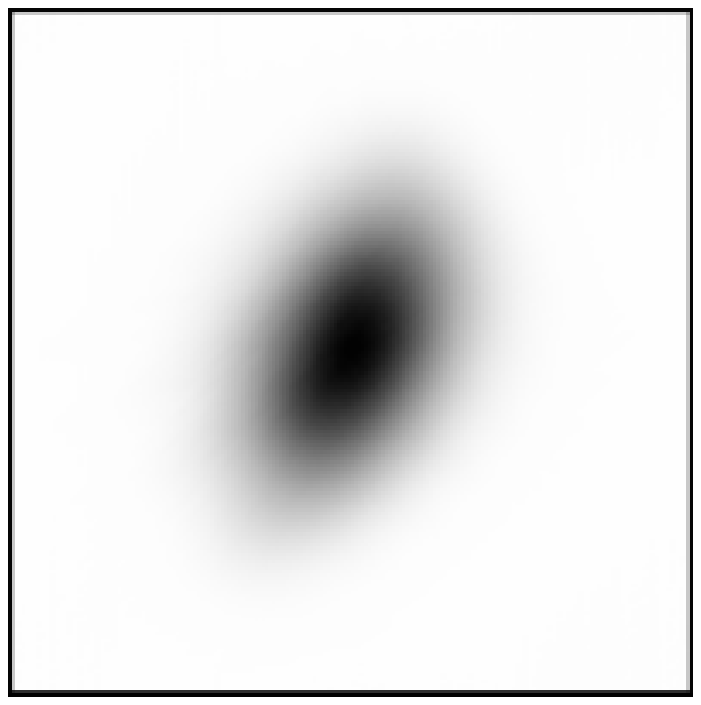} \hspace{0.5cm}
 \includegraphics[angle=0,scale=0.4]{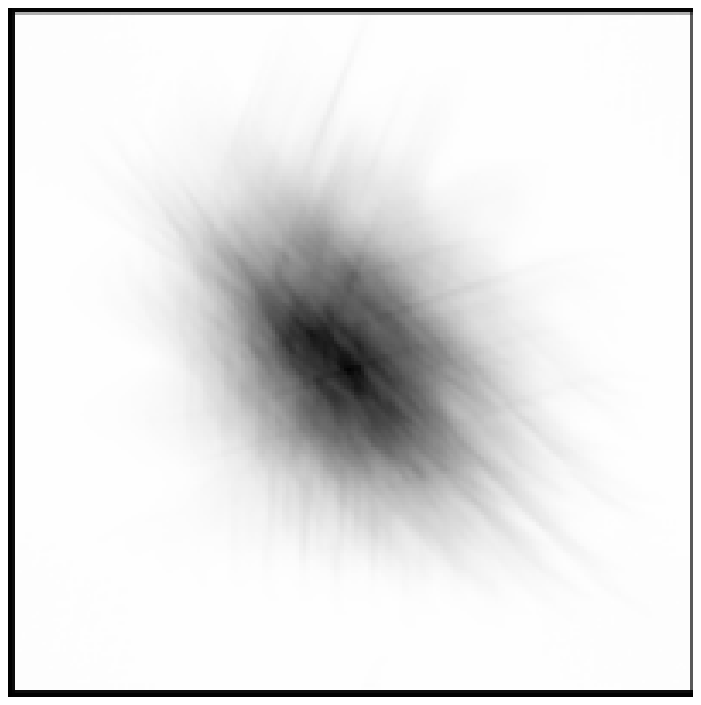} \\
 \vspace{0.2cm} (e) \hspace{2.5cm} (f)
 \caption{Two real neuronal cells (a-b) and their respective total number of
 connections of length $k=2$ (c-d) and 3 (e-f).  The axon has been placed at
 the cell centroid (considering soma plus dendrites).  The neuronal cell
 figures in (a) and (b) are adapted with permission from
 \cite{Wassle:1974}.~\label{fig:ex1}} \end{center}
\end{figure}

\begin{figure}
\begin{center} \vspace{0.3cm}
 \includegraphics[angle=-90,scale=0.4]{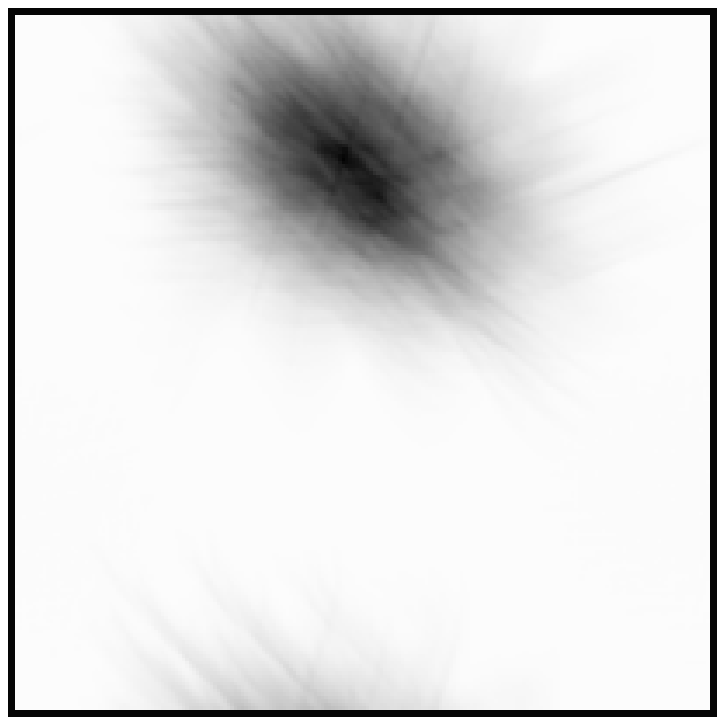} \hspace{0.1cm}
 \includegraphics[angle=-90,scale=0.4]{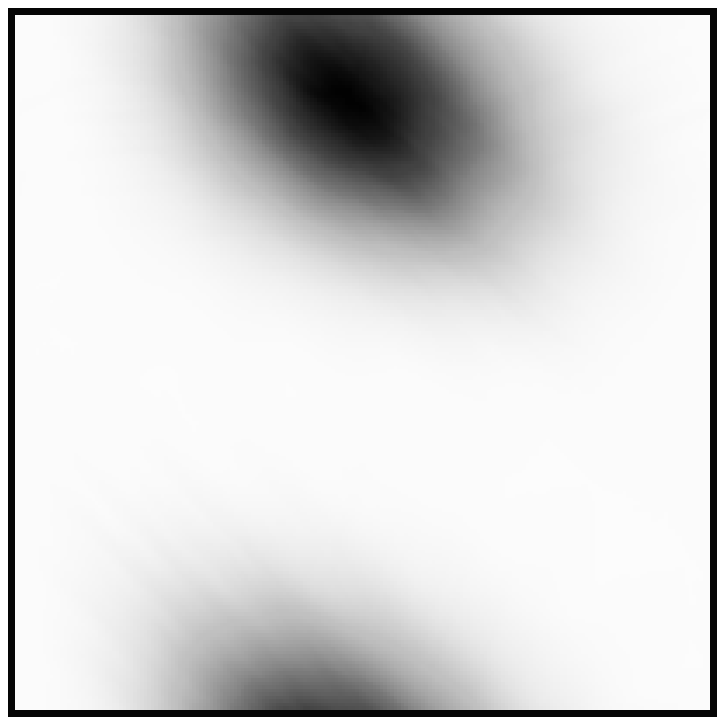} \\
 \vspace{0.5cm} (a) \hspace{2.5cm} (b) \\
 \includegraphics[angle=-90,scale=0.4]{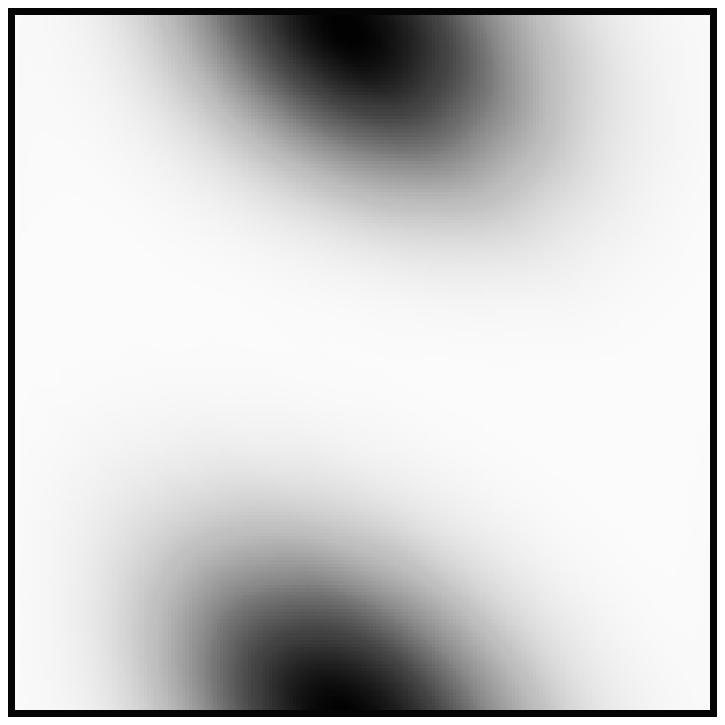} \hspace{0.1cm}
 \includegraphics[angle=-90,scale=0.4]{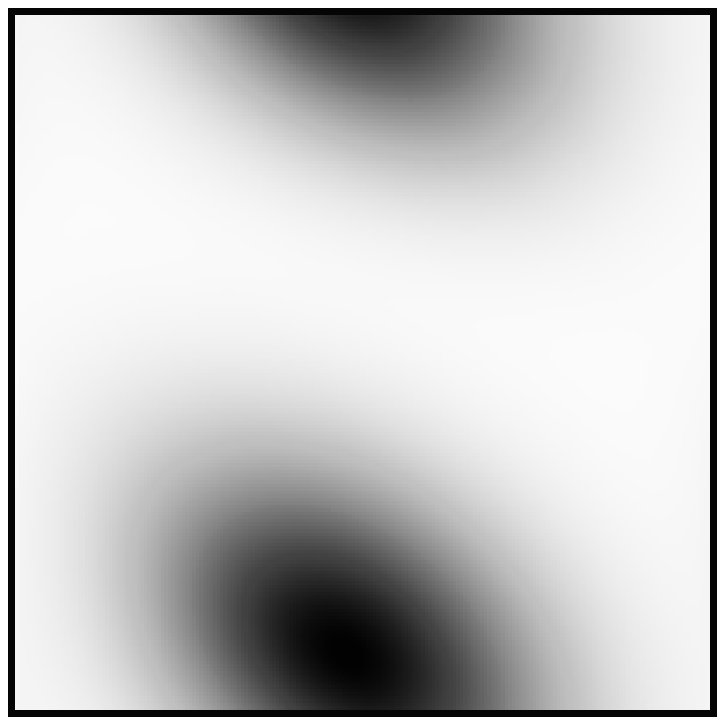} \\
 \vspace{0.5cm} (c) \hspace{2.5cm} (d)
\caption{The total number of connections of length $k=1$ (a), 2 (b), 3 (c)
 and 4(d) for the neuronal cell in Figure~\ref{fig:ex1}(a) with the axon
 placed over the cell centroid, which is itself displaced from the cell
 centroid by $ \vec{s}=(0,7)$ pixels.~\label{fig:ex2}}
 \end{center}
\end{figure}

Figure~\ref{fig:ex1} shows two digital images obtained from real
ganglion cells, (a) and (b), and their respective functions
$\chi_k(x,y)$ for $k=2$ and $3$.  The axon has been placed over the
centroid of the neuronal shape (including soma and dendrites), whereas
the dendritic trees have been spatially sampled into 2033 and 671
pixels, respectively.  Figure~\ref{fig:ex2} shows $\chi_k(x,y)$
obtained for the cell in Figure~\ref{fig:ex1}(a) but with the soma
located at the cell center of mass, which is displaced from the cell
centroid by $\vec{s}=(0,7)$ pixels.  It is clear from such results
that the neuron morphology strongly determines the connectivity
between cells in two important senses: (i) the spatial scattering of
the dendrite points influences the connectivity distribution and (ii)
the relative position of the axon defines how the centroid of the
connections shifts for increasing values of $k$.  While the increased
number of synaptic connections implied by denser neuronal shapes is as
expected, it is clear from the example in Figure~\ref{fig:ex2} that
the distance from axon to cell center of mass implies spatial
displacement of the connection pattern.  Given the predominantly
two-dimensional structure of the mammals' cortex, such an effect
provides an interesting means to transmit information horizontally
along such structures.  In other words, faster signal transmission
along the cortical surface is achieved whenever the axon is placed
further away from the dendritic tree.  As the proper characterization,
classification, analysis and simulation of neuromorphic networks are
all affected by these two interesting phenomena, it is important to
derive objective related measurements.  The next section addresses the
characterization of the morphology of such networks.

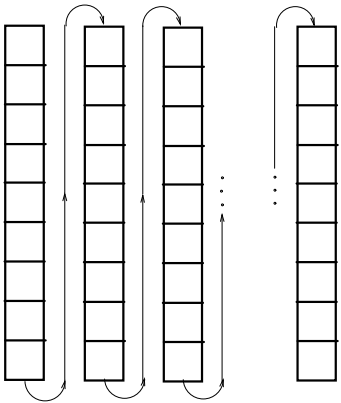
\begin{figure}
\begin{center}
\input{circula.eepic}
  \caption{The circulating scheme. \label{fig:circula}}
\end{center}
\end{figure}
\section{Morphology}

Let $P(x,y)$ be a density function obtained by normalizing
$\chi_k(x,y)$, i.e.

\begin{equation}
  P(x,y)=\frac{\chi(x,y)}{\int_{-\infty}^{\infty}\chi(x,y)dxdy}
\end{equation}

Thus, the spatial scattering of the connections can be quantified in
terms of the respective covariance matrix $K_k$, and the spatial
displacement of the centroid of $P(x,y)$ can be quantified in terms of
the `speed' $v=|| \vec{s} ||$.  Additional geometrical measurements of
the evolution of the neuronal connectivity that can be derived from
the covariance matrix $K$ include the angle $\alpha_k$ that the
distribution main axis makes with the x-axis and the ratio $\rho_k$
between the largest and smallest respective eigenvalues.

Another interesting network feature related to connectivity is its
number $C_{\ell,k}$ of cycles of length $\ell$ established by the
synaptic connections.  This feature can be calculated from the
enlarged matrix $A$ obtained by stacking the columns of the matrix
where the neuronal cell image is represented in order to obtain the
rows of $A$, while the reference point of the cell is shifted along
the main diagonal of $A$.  Figure~\ref{fig:circula} presents the
boundary conditions adopted for the orthogonal lattice $\Omega$
underlying the neuronal structure.  The first line of matrix $A$
corresponds to the whole structure in this figure, starting from the
upper left-most cell and following the arrows.  The second line of $A$
is obtained similarly, but the structure is mounted circularly from
the element (2,2) of $A$, and so on.  As a consequence of such
assumptions, $A$ becomes a circulant matrix.  Observe that studies
involving non-periodical boundary conditions can be approximated by
using large image sizes $N \times N$.

The $N^2$ eigenvalues of the thus obtained \emph{adjacency matrix}
\cite{Albert_Barab:2002} of the whole two-dimensional network are
henceforth represented as $\lambda_i$, $i = 1, 2, \ldots, N^2$.  As
$A$ is circulant, these eigenvalues can be immediately obtained from
the Fourier transform of its first row.  Observe that the simplicity
and speed of such an approach allow for systematic investigation of a
variety of different neuronal shapes.  As the cell reference point is
assumed never to coincide with a dendrite point, we also have that
$\sum_{r=1}^{N} \lambda_{r} = 0$. As $A$ is a non-negative matrix,
there will always be a non-negative eigenvalue $\lambda_M$, called the
\emph{dominant eigenvalue of A}, such that $\lambda_{r} \leq
\lambda_M$ for any $r=1, 2, \ldots, N$.  The \emph{spectral density}
(e.g. \cite{Albert_Barab:2002}) of the adjacency matrix, defined in
Equation~\ref{eq:spec_dens}, where $\lambda_p$ is the $p-$th
eigenvalue of $A$, provides an additional way to characterize the
topology of the obtained networks.

\begin{eqnarray}
 \rho(\lambda)=\frac{1}{N} \sum_{r=1}^{N} \delta(\lambda-\lambda_r)
 \label{eq:spec_dens}
\end{eqnarray}

\begin{figure}
\begin{center}
  \includegraphics[scale=.43,angle=-90]{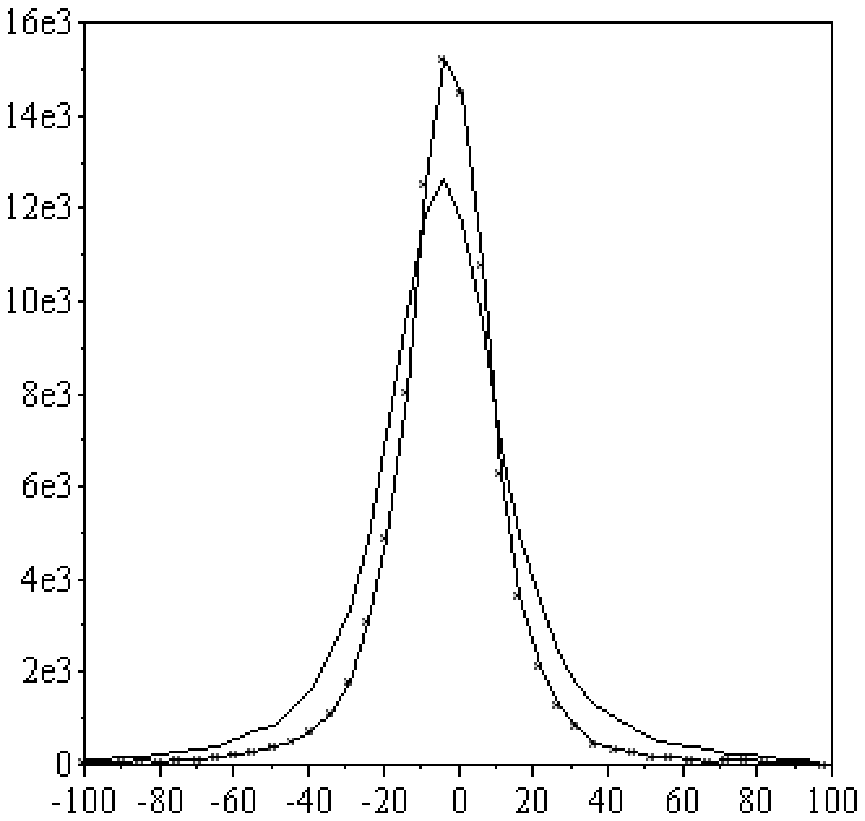} \hspace{0.2cm}
  \includegraphics[scale=.43,angle=-90]{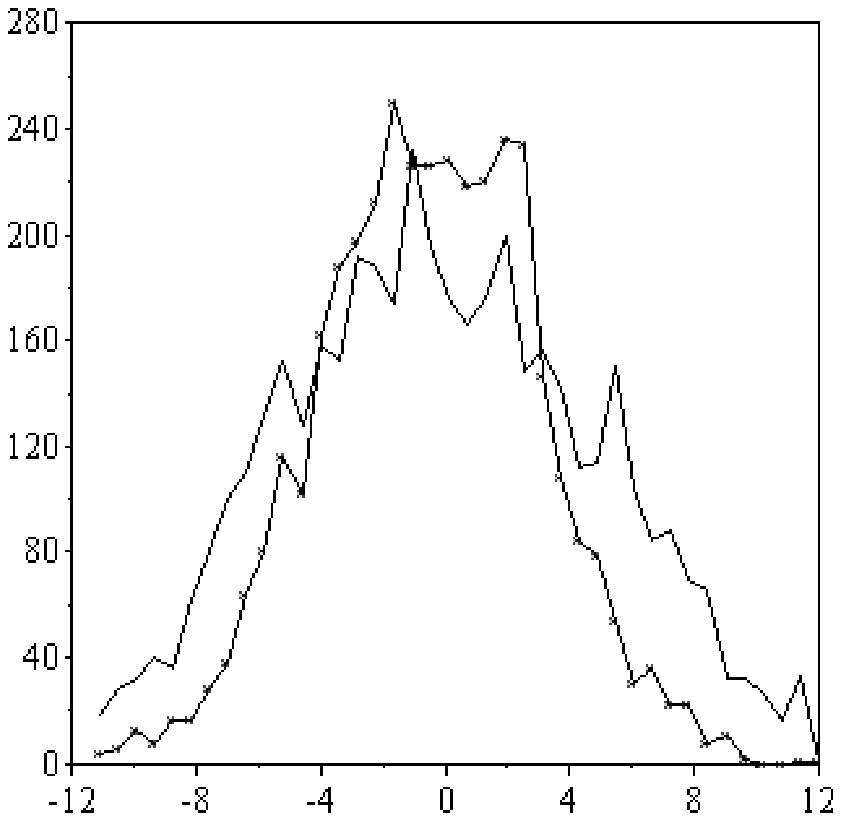} \\
  \vspace{0.2cm} (a) \hspace{4cm} (b)
  \caption{Spectral density of the adjacency matrices obtained for
  the neuronal cells in Figure~\ref{fig:ex1} considering $\Delta=1$
  (a) and $\Delta=5$ (b).  The crossed lines referes to the
  sparser neuronal cell.~\label{fig:eigs}} \end{center}
\end{figure}

The eigenvalue $\lambda_M$, which depends on the specific dynamics
through which new edges are incorporated into the network, represents
an interesting parameter for characterizing the cyclic composition of
complex networks.  Figures~\ref{fig:eigs}(a) and (b) show the real
part (recall that the adjacency matrix for a digraph is not
necessarily symmetric) of the spectral density of the adjacency
matrices obtained for the neuronal cells in Figure~\ref{fig:ex1}(a)
and (b) considering lattice spacings $\Delta=1$ and 5.  The wider
dispersion of the spectrum of the denser cell in
Figure~\ref{fig:eigs}(a) reflects a higher potential for connections
of that neuron in both cases.  It is also clear that the separation of
cells by $\Delta=5$ leads to a substantially smaller spectrum, with
immediate implications for the respective neuronal connectivity.

\begin{figure}
\begin{center}
  \includegraphics[scale=.42,angle=-90]{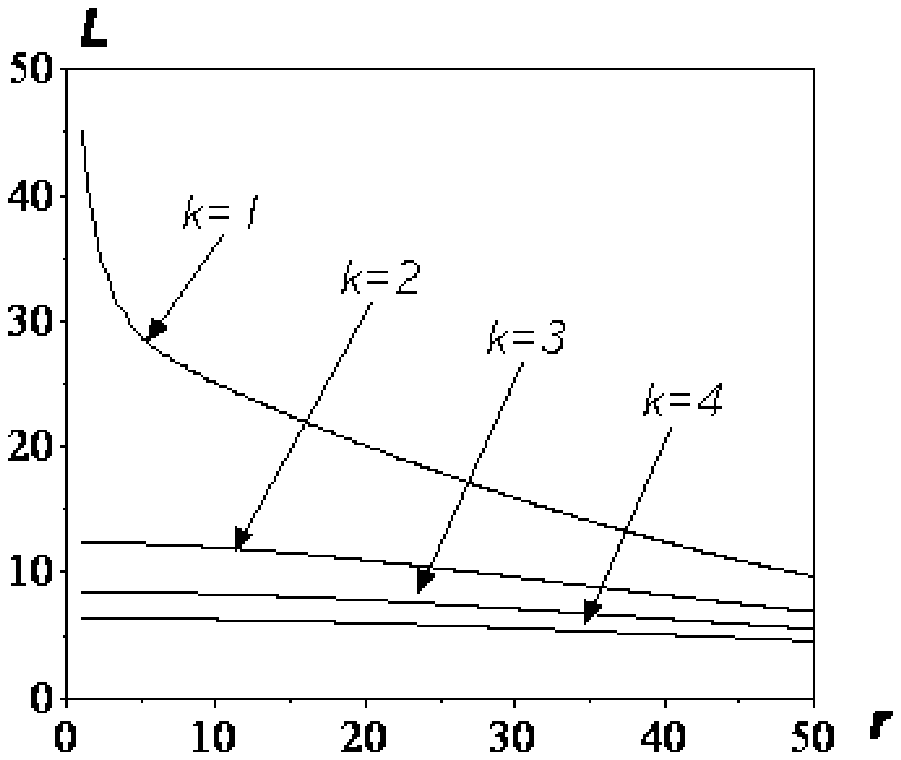}
  \includegraphics[scale=.42,angle=-90]{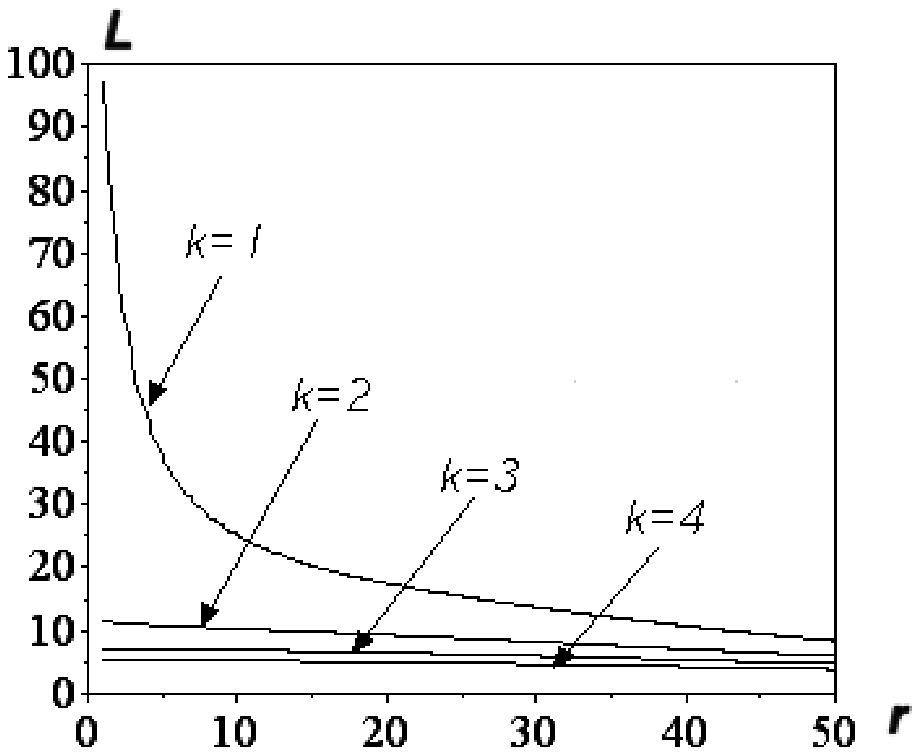} \\
  \vspace{0.2cm} (a) \hspace{4cm} (b)
  \caption{Lacunarities for the spatial distributions of connections for the
  cells in Figure~\ref{fig:ex1}.~\label{fig:lacuns}} \end{center}
\end{figure}

An additional morphological property of the spatial distribution of the
connections is their respective \emph{lacunarity} (e.g. \cite{Hovi:2003}),
which expresses the degree of translational invariance of the obtained
densities.  Figure~\ref{fig:lacuns} shows the lacunarities of the connection
densities obtained for the two considered cells with respect to $k=1$ to $4$.
It is interesting to observe that most of the lacunarity differences are
observed for $k=1$, with similar curves being obtained for larger values of
$k$.  At the same time, the denser cell led to lower lacunarity values.  Given
their immediate implications for neuronal connectivity, the above proposed set
of neuronal shape measurements present specially good potential for neuron
characterization and classification.

\section{Associative Recall}

One of the most important functional properties of Hopfield neuronal networks
is their associative recall, which can be quantified by the overlap
measurement, obtained by comparing the originally trained and the recovered
patterns.  A Hopfield network is completely determined by its respective
connecting matrix, which in the case of the models considered in the present
work is constrained by the adjacency matrix derived from the morphological
structure in the sense that only the weights corresponding to existing
connections are allowed to vary~\cite{Costa_BM:2003}.  Therefore, the shape of
the neuron has a direct impact on the performance of the network, which is
chiefly dictated by the null-space of the adjacency matrix
spectrum~\cite{haykin:1999}.  The analytical approach introduced in the
current work can be immediately used for the estimation of overlaps
considering different neuronal shapes.

In the standard Hopfield setup the cells are either firing ($S_i =1$) or
silent ($S_i =-1$) and are updated according to the rule

\begin{equation}
   S_i \rightarrow {\rm sign} (\sum_k J_{ik} S_k)
\end{equation}

with synaptic strengths $J_{ik} = \sum_\mu \xi_i^\mu \xi_k^\mu$ (Hebb
rule) if $i$ and $k$ are connected, where $\xi_i^\mu = \pm 1, \; \mu =
1,2, \dots P,$ are $P$ random bit-strings called input patterns, and
one of them, perturbed uniformly along its extent, is supposed to be
recalled by this updating rule (i.e. associative memory). The quality
of recall is measured by the overlap $\Psi = \sum_i S_i \xi^1_i$ if the
first pattern is supposed to be recovered. 

\begin{figure}
\begin{center}
\begin{tabular}{lll}
\includegraphics[scale=.15,angle=-90]{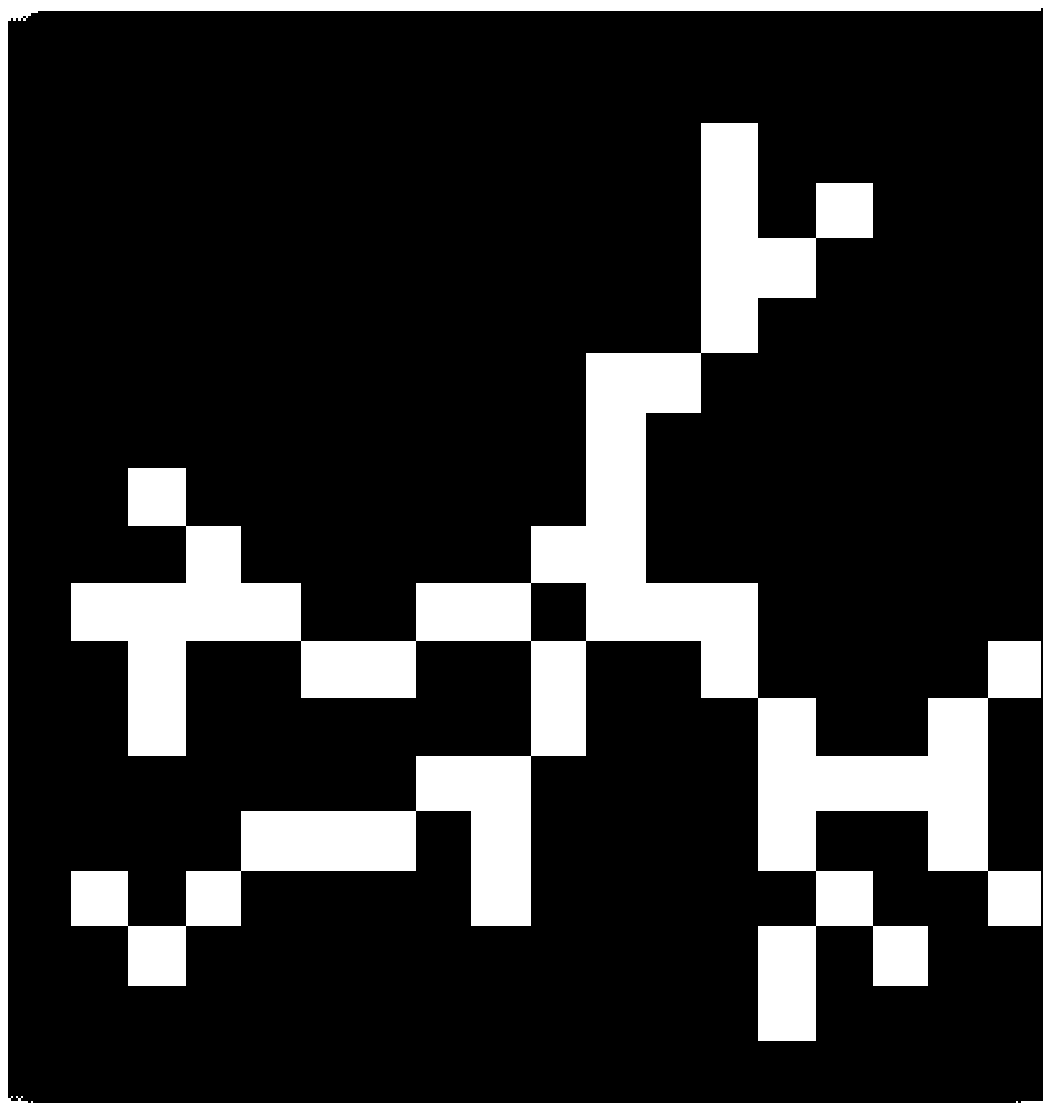}&
\includegraphics[scale=.15,angle=-90]{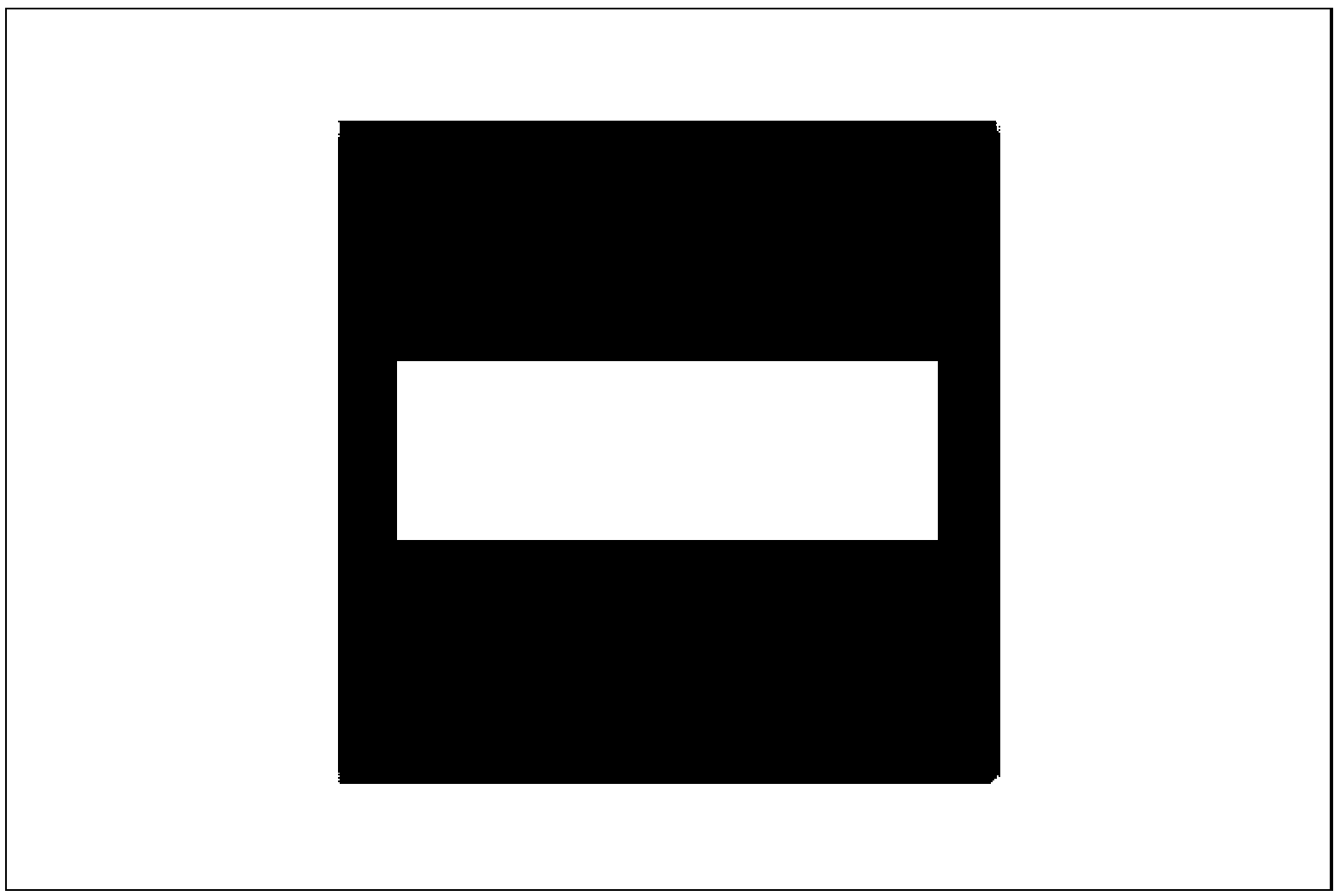}&
\includegraphics[scale=.15,angle=-90]{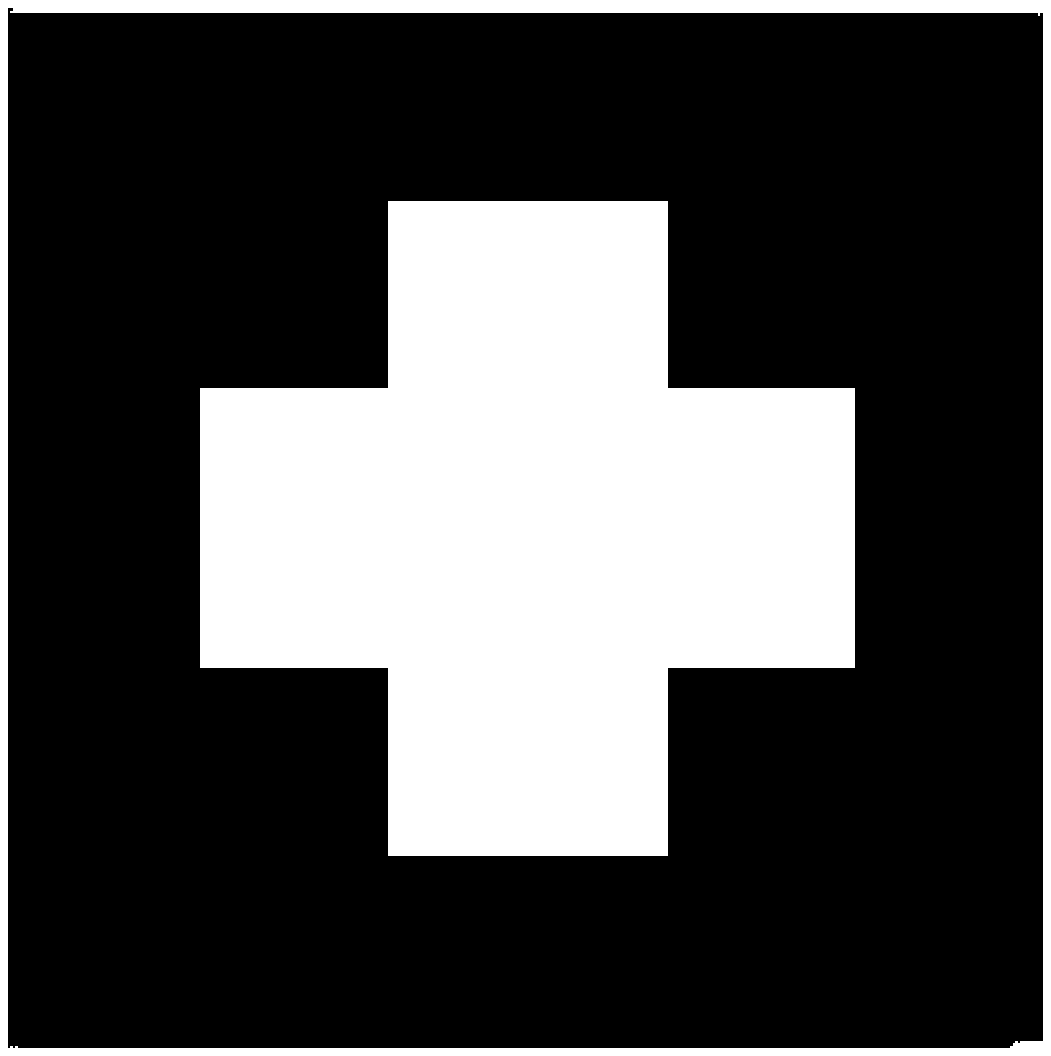}
\end{tabular}
\caption{Three prototypical cell shapes used in the morphological Hopfield
  simulation.}  
\label{fig:try} 
\end{center}
\end{figure}

We consider the simple prototypical cell patterns presented in
Figure~\ref{fig:try}.  A network was obtained for each cell shape by
using the above formalism, and a non-randomly diluted Hopfield model
with such connection matrix was then implemented for the networks and
serially repeated a hundered times to gain statistical
significance. All three networks consists of 441 neurons and the
memory model is set to recall 25 background patterns with 20 percent
of noise. For efficiency's sake, the recovering stage stops whenever a
stable point (or reasonable fixed limit) is reached.  To incorporate
the effect of changing the axons position, which is potential
significant from the biological point of view, we disturbed the
reference point of the axon random and gradually.  This allowed us to
study the robusteness of the memory model.

\begin{figure}
\begin{center}
\includegraphics[scale=.31,angle=-90]{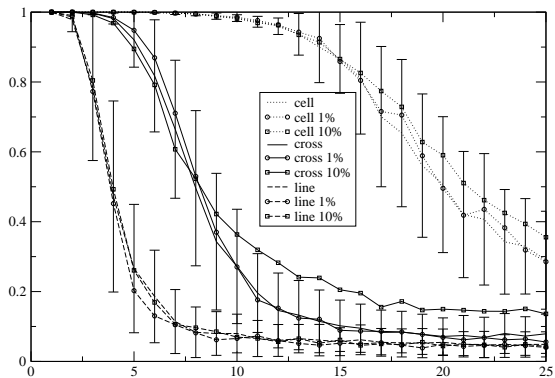}
\caption{The overlap curves for three diferent cell shapes shown in
  Figure~\ref{fig:try}.}  
\label{fig:allovr} 
\end{center}
\end{figure}

\begin{figure}
\begin{center}
\includegraphics[scale=.31,angle=-90]{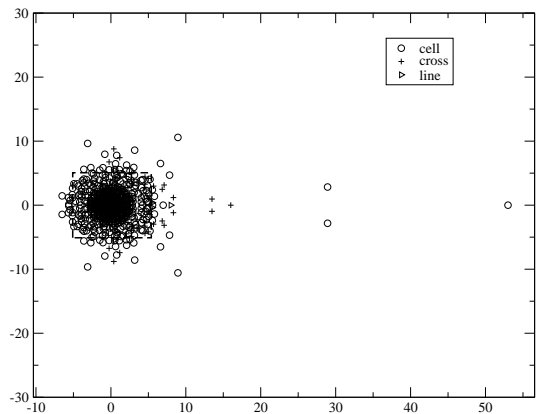} 
\caption{The spectrum of the A matrix (represented in the complex plane) for
  the three diferent cell shapes shown in Figure~\ref{fig:try}, the marked
  inset is zoomed in Figure~\ref{fig:zoomed}. }
\label{fig:spec} 
\end{center}
\end{figure}

\begin{figure}
\begin{center}
\includegraphics[scale=.31,angle=-90]{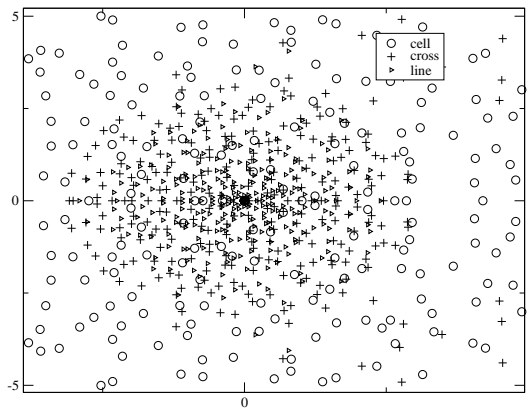} 
\caption{The spectrum of the A matrix zoomed from the inset of
  Figure~\ref{fig:spec}, showing the spread in the distribution of
  eigenvalues for the considered patterns.}  
\label{fig:zoomed} 
\end{center}
\end{figure}

Figure~\ref{fig:allovr} shows the overlaps obtained while considering three
simple neuronal shapes, namely the artificial neuronal cell shown in
Figure~\ref{fig:try}, a line and a cross.  For each cell pattern we have three
overlap curves representing respectively: the lattice as it is, a perturbed
version with one percent of uniform noise (in pixels), and another perturbed
version with ten percent noise. Although the influence of such pertubations
are masked by the stochastic nature of the system, we can see that as far the
overlaps are taken as an indication of performance, the model is robust to
network topology changes and sensitive to neuronal shape. An order clearly
emmerged in the overlaps albeit, as indicated by the large deviation, it is a
visibly degenerated morphological measure.  The more spattially distributed
shapes tended to lead to better preformance.  The eigenvalues for each of the
three considered cases are illustrated graphically in
Figure~\ref{fig:spec}. It is clear from this figure that the eigenvalues tend
to organize symmetrically with respect to the real axis. Also, the eigenvalue
dispersions obtained for the cases line, cross and cell tended to be
progressively broader, reflecting their original spatial structure.  In other
words, shapes more uniform and isotropically distributed along space tended to
produce broader eigenvalue distributions.  The better overlap figures obtained
for sparser patterns is closely related to the fact that more distributed
eigenvalues are obtained in those cases, reducing the null space for memory
representation.

\section{Conclusions}

An analytical approach to the characterization of neuronal
connectivity was proposed and illustrated with respect to synthetic
and real neurons.  It was shown that the connections of progressive
length established along the neuronal structure can be precisely
quantified in terms of the autoconvolution of the function
representing the individual shape of the neuron.  The analysis of the
connectivity pattern in terms of spectral dispersion and lacunarity
was also proposed and illustrated.  In addition, we have shown that,
by assuming a specific type of periodical boundary condition, it is
possible to construct a circulant matrix whose spectrum can be
conveniently calculated and used to characterize the topological and
functional properties of the neuronal structure.  More specifically,
we implemented Hopfield networks having weight matrices constrained by
the adjacency matrices defined by the individual neuronal cell
geometry.  The performance of such networks, quantified in terms of
the overlap measurement, were shown to be strongly related to the
morphology of the adopted neurons.

In addition to paving the way for the analytical characterization of
the connections in regular neuromorphic networks, the framework
proposed in this article can be immediately adapted to study the
spread of waves of neuronal \emph{activity} starting from the stimulus
$\xi(x,y)$, where the neurons are understood to accept input and
produce output in synchronous manner at each clock cycle $T$.  For
instance, the situation where the neuronal cells output corresponds to
the inner product between its shape and the respective area of the
input space can be immediately characterized in terms of the
eigenvalues and eigenvectors of precisely the same adjacency matrix
$A$ constructed as described above.  Although the proposed methodology
assumes identical, uniformly distributed neuronal cells, it is
expected that they provide a reference model for investigating and
characterizing real networks characterized by a certain degree of
regularity, such as some subsystems found in the retina and cortex.
Preliminary corroborations of such a possibility were provided by the
fact that our overlap simulations tended to be robust to lattice
perturbations.  Mean-field extensions of the reported approach are
currently being validated.

\begin{acknowledgments}

The authors are grateful to FAPESP (processes 99/12765-2, 96/05497-3
and 02/02504-01)  and CNPq for financial support.

\end{acknowledgments}


\bibliography{graphm}

\end{document}

%% file: prop.eepic
\setlength{\unitlength}{0.00043745in}
\begingroup\makeatletter\ifx\SetFigFont\undefined%
\gdef\SetFigFont#1#2#3#4#5{%
  \reset@font\fontsize{#1}{#2pt}%
  \fontfamily{#3}\fontseries{#4}\fontshape{#5}%
  \selectfont}%
\fi\endgroup%
{\renewcommand{\dashlinestretch}{30}
\begin{picture}(5469,5338)(0,-10)
\path(2591,1729)(2598,1729)(2604,1729)
	(2617,1722)(2638,1722)(2651,1715)
	(2664,1715)(2678,1715)(2691,1715)
	(2704,1715)(2712,1715)(2719,1715)
	(2726,1715)(2726,1708)(2732,1708)
	(2732,1702)(2739,1702)(2746,1702)
	(2746,1695)(2752,1695)(2760,1695)
	(2760,1688)(2766,1688)(2773,1688)
	(2773,1681)(2779,1681)(2779,1688)
	(2779,1702)(2786,1708)(2786,1722)
	(2792,1735)(2792,1742)(2792,1755)
	(2792,1762)(2792,1776)(2792,1782)
	(2792,1795)(2792,1802)(2792,1816)
	(2792,1823)(2792,1837)(2792,1843)
	(2786,1857)(2786,1864)(2786,1870)
	(2786,1877)(2786,1870)(2786,1864)
	(2786,1857)(2786,1850)(2786,1837)
	(2786,1823)(2786,1809)(2786,1795)
	(2786,1782)(2786,1769)(2786,1762)
	(2786,1755)(2786,1749)(2786,1742)
	(2786,1735)(2786,1729)(2786,1722)
	(2786,1715)(2786,1708)(2786,1702)
	(2786,1695)(2786,1688)(2786,1681)
\path(2786,1681)(2792,1674)(2792,1667)
	(2800,1667)(2806,1654)(2813,1654)
	(2820,1641)(2826,1641)(2826,1634)
	(2833,1627)(2840,1620)(2840,1614)
	(2848,1607)(2854,1601)(2861,1601)
	(2861,1594)(2867,1587)(2874,1580)
	(2880,1580)(2880,1573)(2888,1573)
	(2895,1567)(2895,1560)(2901,1560)
	(2908,1554)(2908,1546)(2914,1546)
	(2914,1539)(2921,1539)(2921,1532)
	(2921,1526)(2928,1526)(2928,1519)
	(2935,1519)(2935,1513)(2941,1506)
	(2948,1506)(2948,1499)
\path(2948,1506)(2948,1519)(2948,1526)
	(2948,1532)(2954,1539)(2954,1546)
	(2954,1554)(2954,1560)(2954,1573)
	(2954,1580)(2954,1587)(2954,1594)
	(2954,1601)(2954,1607)(2954,1614)
	(2954,1620)(2954,1627)(2954,1634)
	(2954,1641)(2954,1647)(2948,1654)
	(2948,1660)(2948,1667)(2948,1674)
	(2948,1681)(2948,1688)(2948,1695)
	(2948,1702)(2948,1708)(2948,1715)
	(2948,1722)(2948,1729)(2954,1729)
	(2954,1735)(2954,1742)(2954,1749)
	(2954,1755)(2954,1762)(2954,1769)
	(2954,1776)(2954,1782)(2954,1789)
	(2954,1802)(2954,1809)(2954,1816)
	(2948,1823)(2948,1829)(2948,1837)
	(2948,1843)(2948,1850)(2948,1857)
	(2948,1864)(2948,1870)(2941,1877)
	(2941,1883)(2941,1890)(2941,1897)
	(2941,1904)(2941,1911)(2941,1917)
	(2935,1924)(2935,1930)(2935,1937)
	(2935,1944)(2935,1951)(2935,1957)(2935,1964)
\path(2948,1513)(2948,1519)(2948,1526)
	(2954,1532)(2961,1539)(2961,1546)
	(2968,1546)(2968,1554)(2976,1560)
	(2983,1567)(2989,1573)(2989,1580)
	(2996,1587)(3002,1587)(3002,1594)
	(3009,1601)(3016,1607)(3016,1614)
	(3023,1614)(3023,1620)(3023,1627)
	(3029,1627)(3029,1634)(3029,1641)
	(3036,1641)(3036,1647)(3036,1654)
	(3036,1660)(3036,1667)(3036,1674)
	(3036,1681)(3036,1688)(3036,1695)
	(3036,1702)(3036,1708)(3036,1715)
	(3036,1722)(3036,1729)(3036,1735)
	(3036,1742)(3036,1749)(3036,1755)
	(3036,1762)(3036,1769)(3036,1776)
	(3036,1782)(3043,1782)(3043,1789)
	(3043,1795)(3043,1802)(3043,1809)
	(3049,1809)(3049,1816)(3049,1823)
	(3049,1829)(3056,1829)(3056,1837)
	(3056,1843)(3063,1843)(3070,1843)
	(3070,1850)(3076,1857)(3083,1857)
	(3083,1864)(3089,1870)(3096,1870)
	(3096,1877)(3104,1877)(3111,1883)
	(3118,1890)(3124,1897)(3131,1897)
	(3131,1904)(3137,1911)(3144,1917)
	(3144,1924)(3151,1924)(3151,1930)
	(3151,1937)(3158,1944)(3158,1951)
	(3158,1957)(3158,1964)(3164,1970)
	(3164,1977)(3164,1985)(3164,1992)
	(3164,1999)(3171,2005)(3171,2012)
	(3171,2018)(3177,2025)(3177,2032)
	(3177,2039)(3177,2045)(3184,2045)(3184,2052)
\path(3131,1883)(3131,1890)(3137,1890)
	(3137,1897)(3144,1897)(3151,1897)
	(3151,1904)(3158,1904)(3164,1904)
	(3164,1911)(3171,1911)(3177,1911)
	(3184,1917)(3192,1917)(3198,1917)
	(3205,1924)(3211,1924)(3218,1924)
	(3224,1924)(3232,1924)(3232,1930)
	(3239,1930)(3246,1930)(3252,1930)(3259,1930)
\path(2948,1519)(2948,1513)(2948,1506)
	(2948,1499)(2941,1492)(2941,1485)
	(2941,1479)(2935,1479)(2935,1472)
	(2935,1466)(2928,1459)(2928,1452)
	(2921,1445)(2921,1438)(2921,1432)(2914,1425)
\path(2928,1452)(2935,1452)(2941,1452)
	(2948,1452)(2954,1452)(2961,1452)
	(2968,1452)(2976,1452)(2983,1452)
	(2989,1452)(2996,1452)(2996,1459)
	(3002,1459)(3009,1452)
\path(3009,1452)(3009,1459)(3009,1466)
	(3016,1472)(3023,1472)(3023,1479)
	(3029,1479)(3036,1485)(3036,1492)
	(3043,1492)(3043,1499)(3049,1506)
	(3049,1513)(3056,1513)(3056,1519)
	(3063,1519)(3063,1526)(3070,1526)
	(3076,1532)(3076,1539)(3083,1539)
	(3089,1546)(3089,1554)(3096,1554)
	(3104,1560)(3104,1567)(3111,1567)
	(3111,1573)(3118,1573)(3118,1580)
	(3124,1580)(3124,1587)(3131,1587)
	(3137,1587)(3137,1594)(3144,1594)
	(3144,1601)(3151,1601)(3158,1607)
	(3164,1607)(3164,1614)(3171,1614)
	(3171,1620)(3177,1627)(3177,1634)
	(3184,1634)(3184,1641)(3192,1641)
	(3192,1647)(3198,1647)(3198,1654)
	(3205,1654)(3205,1660)(3211,1660)
	(3211,1667)(3218,1667)(3224,1667)
	(3232,1667)(3239,1660)(3246,1660)
	(3246,1667)(3246,1674)(3246,1681)
	(3252,1688)(3252,1695)(3252,1702)
	(3259,1708)(3259,1715)(3259,1722)
	(3266,1729)(3266,1735)(3266,1742)
	(3272,1742)(3272,1749)(3280,1755)
	(3280,1762)(3286,1769)(3293,1776)
	(3293,1782)(3299,1789)(3306,1795)
	(3306,1802)(3312,1802)(3312,1809)
	(3320,1809)(3320,1816)(3320,1823)
	(3326,1823)(3333,1829)(3333,1837)
	(3340,1837)(3346,1837)(3346,1843)
	(3353,1843)(3360,1843)(3360,1850)
	(3368,1850)(3374,1850)(3374,1857)
	(3381,1857)(3387,1857)(3387,1864)
	(3394,1864)(3394,1870)(3400,1870)
	(3408,1870)(3408,1877)(3415,1877)
	(3415,1883)(3421,1890)(3428,1890)
	(3428,1897)(3434,1897)(3434,1904)
	(3441,1904)(3448,1911)(3455,1911)
	(3461,1917)(3468,1917)(3474,1924)
	(3481,1924)(3488,1930)(3496,1930)
	(3503,1930)(3503,1937)(3503,1944)
	(3503,1951)(3509,1957)(3509,1964)
	(3516,1970)(3516,1977)(3522,1977)
	(3529,1985)(3536,1985)(3543,1985)
	(3543,1992)(3549,1992)(3556,1992)(3556,1999)
\path(3158,1594)(3158,1601)(3158,1607)
	(3164,1607)(3171,1607)(3171,1614)
	(3177,1614)(3184,1620)(3192,1620)
	(3198,1627)(3205,1627)(3211,1627)
	(3218,1634)(3224,1634)(3232,1634)
	(3239,1634)(3246,1634)(3246,1641)
	(3252,1641)(3259,1647)(3266,1647)
	(3266,1654)(3272,1654)(3272,1660)
	(3280,1660)(3280,1667)(3280,1674)
	(3286,1674)(3286,1681)(3293,1688)
	(3293,1695)(3293,1702)(3299,1702)
	(3299,1708)(3299,1715)(3306,1722)
	(3306,1729)(3312,1729)(3312,1735)
	(3312,1742)(3320,1742)(3320,1749)
	(3326,1749)(3326,1755)(3333,1755)
\path(3340,1357)(3346,1364)(3353,1371)
	(3360,1378)(3368,1384)(3374,1384)
	(3381,1391)(3387,1391)(3394,1397)
	(3400,1397)(3408,1397)(3408,1405)
	(3415,1405)(3421,1412)(3428,1419)
	(3434,1419)(3434,1425)(3441,1425)
	(3448,1425)(3448,1432)(3448,1438)(3455,1445)
\path(3198,1391)(3198,1384)(3198,1378)
	(3198,1371)(3198,1364)(3205,1364)
	(3205,1357)(3211,1357)(3211,1350)
\path(2921,1425)(2921,1419)(2921,1412)
	(2921,1405)(2921,1397)(2921,1391)
	(2921,1384)(2921,1378)(2921,1371)
	(2921,1364)(2921,1357)(2921,1350)
	(2921,1344)(2921,1337)(2921,1331)
	(2921,1324)(2921,1317)(2921,1310)
	(2921,1304)(2921,1297)(2921,1291)(2921,1284)
\path(2921,1284)(2928,1284)(2935,1284)
	(2935,1277)(2941,1277)(2941,1270)
	(2948,1270)(2948,1263)(2948,1256)
	(2954,1256)(2954,1249)(2961,1249)
	(2961,1243)(2961,1236)(2968,1236)
	(2968,1229)(2976,1229)(2976,1222)
	(2976,1216)(2983,1216)(2989,1216)
	(2989,1209)(2989,1202)(2996,1202)
	(2996,1196)(3002,1196)(3002,1189)
	(3002,1182)(3002,1175)(3002,1169)
	(3002,1162)(3002,1156)(3002,1149)
	(3002,1142)(3002,1135)(3002,1128)
	(3002,1122)(3002,1114)(3002,1108)
	(3002,1101)(3002,1094)(3002,1087)
	(3002,1081)(3002,1074)(3002,1068)
	(3002,1061)(3009,1061)(3009,1054)
	(3009,1047)(3016,1047)(3016,1040)
	(3023,1040)(3023,1034)(3029,1034)
	(3029,1027)(3036,1027)(3043,1027)
	(3043,1021)(3043,1014)(3049,1014)
	(3049,1007)(3056,1007)(3056,1000)
	(3056,994)(3063,994)(3063,987)
	(3063,981)(3070,974)(3070,966)
	(3070,959)(3076,959)(3076,952)
	(3076,946)(3083,939)(3083,933)
	(3083,926)(3083,919)(3083,912)
	(3089,912)(3089,906)(3089,899)
\path(2928,1452)(2921,1452)(2914,1452)
	(2908,1445)(2901,1445)(2895,1438)
	(2888,1438)(2888,1432)(2880,1432)
	(2874,1432)(2874,1425)(2867,1425)
	(2861,1425)(2861,1419)(2854,1419)
	(2854,1412)(2848,1412)(2848,1405)
	(2848,1397)(2840,1397)(2840,1391)
	(2840,1384)(2840,1378)(2833,1371)
	(2833,1364)(2833,1357)(2833,1350)
	(2833,1344)(2833,1337)(2833,1331)
	(2833,1324)(2833,1317)(2833,1310)
	(2833,1304)(2833,1297)(2833,1291)
	(2826,1291)(2826,1284)(2820,1277)
	(2820,1270)(2813,1263)(2813,1256)
	(2806,1249)(2806,1243)(2800,1236)
	(2800,1229)(2800,1222)(2792,1222)
	(2792,1216)(2786,1209)(2786,1202)
	(2786,1196)(2779,1196)(2779,1189)
	(2779,1182)(2779,1175)(2779,1169)
	(2773,1169)(2773,1162)(2773,1156)
	(2766,1156)(2766,1149)(2766,1142)
	(2766,1135)(2766,1128)(2760,1122)
	(2760,1114)(2752,1108)(2752,1101)
	(2752,1094)(2746,1087)(2739,1081)
	(2739,1074)(2732,1074)(2732,1068)
	(2726,1068)(2726,1061)(2719,1061)
	(2719,1054)(2719,1047)(2719,1040)
	(2712,1040)(2712,1034)(2712,1027)
\path(2928,1459)(2928,1452)(2921,1452)
	(2914,1452)(2908,1452)(2901,1445)
	(2895,1445)(2888,1445)(2880,1438)
	(2874,1438)(2861,1438)(2854,1438)
	(2848,1438)(2840,1432)(2833,1432)
	(2826,1425)(2820,1425)(2813,1419)
	(2806,1419)(2806,1412)(2800,1412)
	(2792,1412)(2786,1405)(2779,1397)
	(2773,1397)(2766,1391)(2760,1391)
	(2752,1384)(2746,1378)(2739,1378)
	(2732,1371)(2726,1364)(2719,1357)
	(2712,1350)(2704,1350)(2704,1344)
	(2698,1344)(2698,1337)(2691,1337)
	(2691,1331)(2685,1331)(2685,1324)
	(2678,1317)(2672,1317)(2672,1310)
	(2664,1310)(2664,1304)(2657,1304)
	(2657,1297)(2657,1291)(2651,1291)
	(2651,1284)(2651,1277)(2644,1277)
	(2644,1270)(2644,1263)(2644,1256)
	(2644,1249)(2638,1243)(2638,1236)
	(2638,1229)(2631,1229)(2631,1222)
	(2631,1216)(2624,1216)(2624,1209)
	(2617,1209)(2617,1202)(2611,1196)
	(2604,1189)(2604,1182)(2598,1175)
	(2598,1169)(2591,1169)(2591,1162)
	(2584,1156)(2584,1149)(2576,1149)
	(2576,1142)(2569,1135)(2569,1128)
	(2563,1128)(2563,1122)(2556,1122)
	(2556,1114)(2550,1114)(2550,1108)
	(2543,1108)(2543,1101)(2536,1101)
	(2536,1094)(2529,1087)(2529,1081)
	(2523,1081)(2523,1074)(2523,1068)
	(2516,1068)(2516,1061)(2516,1054)
	(2516,1047)(2516,1040)(2516,1034)
	(2516,1027)(2516,1021)(2516,1014)
	(2516,1007)(2516,1000)(2516,994)
	(2516,987)(2516,981)(2516,974)(2516,966)
\path(3144,1438)(3137,1438)(3131,1438)
	(3124,1438)(3118,1438)(3111,1438)
	(3104,1438)(3096,1438)(3089,1438)
	(3083,1438)(3076,1438)
\path(3002,1459)(3009,1459)(3009,1466)
	(3016,1466)(3023,1466)(3036,1466)
	(3043,1466)(3049,1466)(3056,1466)
	(3063,1466)(3070,1466)(3076,1466)
	(3083,1466)(3089,1466)(3089,1459)
	(3096,1459)(3096,1452)(3104,1452)
	(3104,1445)(3111,1445)(3111,1438)
	(3118,1438)(3118,1432)(3124,1432)
	(3124,1425)(3131,1425)(3131,1419)
	(3137,1419)(3144,1412)(3151,1412)
	(3158,1412)(3164,1412)(3164,1405)
	(3171,1405)(3177,1405)(3184,1397)
	(3192,1397)(3198,1397)(3205,1397)
	(3205,1391)(3211,1391)(3218,1391)
	(3224,1391)(3232,1391)(3239,1391)
	(3239,1384)(3246,1384)(3252,1384)
	(3259,1378)(3266,1378)(3266,1371)
	(3272,1371)(3280,1371)(3280,1364)
	(3286,1364)(3286,1357)(3293,1357)
	(3293,1350)(3299,1350)(3306,1344)
	(3312,1344)(3320,1344)(3326,1344)
	(3333,1344)(3333,1337)(3340,1337)
	(3346,1337)(3346,1331)(3353,1331)
	(3360,1331)(3368,1331)(3374,1331)
	(3381,1331)(3387,1331)(3394,1331)
	(3400,1331)(3408,1331)(3415,1331)
	(3421,1331)(3428,1331)(3434,1331)
	(3441,1331)(3448,1331)(3455,1331)
	(3461,1331)(3468,1331)
\path(3728,3954)(3735,3954)(3741,3954)
	(3754,3947)(3775,3947)(3788,3940)
	(3801,3940)(3815,3940)(3828,3940)
	(3841,3940)(3849,3940)(3856,3940)
	(3863,3940)(3863,3933)(3869,3933)
	(3869,3927)(3876,3927)(3883,3927)
	(3883,3920)(3889,3920)(3897,3920)
	(3897,3913)(3903,3913)(3910,3913)
	(3910,3906)(3916,3906)(3916,3913)
	(3916,3927)(3923,3933)(3923,3947)
	(3929,3960)(3929,3967)(3929,3980)
	(3929,3987)(3929,4001)(3929,4007)
	(3929,4020)(3929,4027)(3929,4041)
	(3929,4048)(3929,4062)(3929,4068)
	(3923,4082)(3923,4089)(3923,4095)
	(3923,4102)(3923,4095)(3923,4089)
	(3923,4082)(3923,4075)(3923,4062)
	(3923,4048)(3923,4034)(3923,4020)
	(3923,4007)(3923,3994)(3923,3987)
	(3923,3980)(3923,3974)(3923,3967)
	(3923,3960)(3923,3954)(3923,3947)
	(3923,3940)(3923,3933)(3923,3927)
	(3923,3920)(3923,3913)(3923,3906)
\path(3923,3906)(3929,3899)(3929,3892)
	(3937,3892)(3943,3879)(3950,3879)
	(3957,3866)(3963,3866)(3963,3859)
	(3970,3852)(3977,3845)(3977,3839)
	(3985,3832)(3991,3826)(3998,3826)
	(3998,3819)(4004,3812)(4011,3805)
	(4017,3805)(4017,3798)(4025,3798)
	(4032,3792)(4032,3785)(4038,3785)
	(4045,3779)(4045,3771)(4051,3771)
	(4051,3764)(4058,3764)(4058,3757)
	(4058,3751)(4065,3751)(4065,3744)
	(4072,3744)(4072,3738)(4078,3731)
	(4085,3731)(4085,3724)
\path(4085,3731)(4085,3744)(4085,3751)
	(4085,3757)(4091,3764)(4091,3771)
	(4091,3779)(4091,3785)(4091,3798)
	(4091,3805)(4091,3812)(4091,3819)
	(4091,3826)(4091,3832)(4091,3839)
	(4091,3845)(4091,3852)(4091,3859)
	(4091,3866)(4091,3872)(4085,3879)
	(4085,3885)(4085,3892)(4085,3899)
	(4085,3906)(4085,3913)(4085,3920)
	(4085,3927)(4085,3933)(4085,3940)
	(4085,3947)(4085,3954)(4091,3954)
	(4091,3960)(4091,3967)(4091,3974)
	(4091,3980)(4091,3987)(4091,3994)
	(4091,4001)(4091,4007)(4091,4014)
	(4091,4027)(4091,4034)(4091,4041)
	(4085,4048)(4085,4054)(4085,4062)
	(4085,4068)(4085,4075)(4085,4082)
	(4085,4089)(4085,4095)(4078,4102)
	(4078,4108)(4078,4115)(4078,4122)
	(4078,4129)(4078,4136)(4078,4142)
	(4072,4149)(4072,4155)(4072,4162)
	(4072,4169)(4072,4176)(4072,4182)(4072,4189)
\path(4085,3738)(4085,3744)(4085,3751)
	(4091,3757)(4098,3764)(4098,3771)
	(4105,3771)(4105,3779)(4113,3785)
	(4120,3792)(4126,3798)(4126,3805)
	(4133,3812)(4139,3812)(4139,3819)
	(4146,3826)(4153,3832)(4153,3839)
	(4160,3839)(4160,3845)(4160,3852)
	(4166,3852)(4166,3859)(4166,3866)
	(4173,3866)(4173,3872)(4173,3879)
	(4173,3885)(4173,3892)(4173,3899)
	(4173,3906)(4173,3913)(4173,3920)
	(4173,3927)(4173,3933)(4173,3940)
	(4173,3947)(4173,3954)(4173,3960)
	(4173,3967)(4173,3974)(4173,3980)
	(4173,3987)(4173,3994)(4173,4001)
	(4173,4007)(4180,4007)(4180,4014)
	(4180,4020)(4180,4027)(4180,4034)
	(4186,4034)(4186,4041)(4186,4048)
	(4186,4054)(4193,4054)(4193,4062)
	(4193,4068)(4200,4068)(4207,4068)
	(4207,4075)(4213,4082)(4220,4082)
	(4220,4089)(4226,4095)(4233,4095)
	(4233,4102)(4241,4102)(4248,4108)
	(4255,4115)(4261,4122)(4268,4122)
	(4268,4129)(4274,4136)(4281,4142)
	(4281,4149)(4288,4149)(4288,4155)
	(4288,4162)(4295,4169)(4295,4176)
	(4295,4182)(4295,4189)(4301,4195)
	(4301,4202)(4301,4210)(4301,4217)
	(4301,4224)(4308,4230)(4308,4237)
	(4308,4243)(4314,4250)(4314,4257)
	(4314,4264)(4314,4270)(4321,4270)(4321,4277)
\path(4268,4108)(4268,4115)(4274,4115)
	(4274,4122)(4281,4122)(4288,4122)
	(4288,4129)(4295,4129)(4301,4129)
	(4301,4136)(4308,4136)(4314,4136)
	(4321,4142)(4329,4142)(4335,4142)
	(4342,4149)(4348,4149)(4355,4149)
	(4361,4149)(4369,4149)(4369,4155)
	(4376,4155)(4383,4155)(4389,4155)(4396,4155)
\path(4085,3744)(4085,3738)(4085,3731)
	(4085,3724)(4078,3717)(4078,3710)
	(4078,3704)(4072,3704)(4072,3697)
	(4072,3691)(4065,3684)(4065,3677)
	(4058,3670)(4058,3663)(4058,3657)(4051,3650)
\path(4065,3677)(4072,3677)(4078,3677)
	(4085,3677)(4091,3677)(4098,3677)
	(4105,3677)(4113,3677)(4120,3677)
	(4126,3677)(4133,3677)(4133,3684)
	(4139,3684)(4146,3677)
\path(4146,3677)(4146,3684)(4146,3691)
	(4153,3697)(4160,3697)(4160,3704)
	(4166,3704)(4173,3710)(4173,3717)
	(4180,3717)(4180,3724)(4186,3731)
	(4186,3738)(4193,3738)(4193,3744)
	(4200,3744)(4200,3751)(4207,3751)
	(4213,3757)(4213,3764)(4220,3764)
	(4226,3771)(4226,3779)(4233,3779)
	(4241,3785)(4241,3792)(4248,3792)
	(4248,3798)(4255,3798)(4255,3805)
	(4261,3805)(4261,3812)(4268,3812)
	(4274,3812)(4274,3819)(4281,3819)
	(4281,3826)(4288,3826)(4295,3832)
	(4301,3832)(4301,3839)(4308,3839)
	(4308,3845)(4314,3852)(4314,3859)
	(4321,3859)(4321,3866)(4329,3866)
	(4329,3872)(4335,3872)(4335,3879)
	(4342,3879)(4342,3885)(4348,3885)
	(4348,3892)(4355,3892)(4361,3892)
	(4369,3892)(4376,3885)(4383,3885)
	(4383,3892)(4383,3899)(4383,3906)
	(4389,3913)(4389,3920)(4389,3927)
	(4396,3933)(4396,3940)(4396,3947)
	(4403,3954)(4403,3960)(4403,3967)
	(4409,3967)(4409,3974)(4417,3980)
	(4417,3987)(4423,3994)(4430,4001)
	(4430,4007)(4436,4014)(4443,4020)
	(4443,4027)(4449,4027)(4449,4034)
	(4457,4034)(4457,4041)(4457,4048)
	(4463,4048)(4470,4054)(4470,4062)
	(4477,4062)(4483,4062)(4483,4068)
	(4490,4068)(4497,4068)(4497,4075)
	(4505,4075)(4511,4075)(4511,4082)
	(4518,4082)(4524,4082)(4524,4089)
	(4531,4089)(4531,4095)(4537,4095)
	(4545,4095)(4545,4102)(4552,4102)
	(4552,4108)(4558,4115)(4565,4115)
	(4565,4122)(4571,4122)(4571,4129)
	(4578,4129)(4585,4136)(4592,4136)
	(4598,4142)(4605,4142)(4611,4149)
	(4618,4149)(4625,4155)(4633,4155)
	(4640,4155)(4640,4162)(4640,4169)
	(4640,4176)(4646,4182)(4646,4189)
	(4653,4195)(4653,4202)(4659,4202)
	(4666,4210)(4673,4210)(4680,4210)
	(4680,4217)(4686,4217)(4693,4217)(4693,4224)
\path(4295,3819)(4295,3826)(4295,3832)
	(4301,3832)(4308,3832)(4308,3839)
	(4314,3839)(4321,3845)(4329,3845)
	(4335,3852)(4342,3852)(4348,3852)
	(4355,3859)(4361,3859)(4369,3859)
	(4376,3859)(4383,3859)(4383,3866)
	(4389,3866)(4396,3872)(4403,3872)
	(4403,3879)(4409,3879)(4409,3885)
	(4417,3885)(4417,3892)(4417,3899)
	(4423,3899)(4423,3906)(4430,3913)
	(4430,3920)(4430,3927)(4436,3927)
	(4436,3933)(4436,3940)(4443,3947)
	(4443,3954)(4449,3954)(4449,3960)
	(4449,3967)(4457,3967)(4457,3974)
	(4463,3974)(4463,3980)(4470,3980)
\path(4477,3582)(4483,3589)(4490,3596)
	(4497,3603)(4505,3609)(4511,3609)
	(4518,3616)(4524,3616)(4531,3622)
	(4537,3622)(4545,3622)(4545,3630)
	(4552,3630)(4558,3637)(4565,3644)
	(4571,3644)(4571,3650)(4578,3650)
	(4585,3650)(4585,3657)(4585,3663)(4592,3670)
\path(4335,3616)(4335,3609)(4335,3603)
	(4335,3596)(4335,3589)(4342,3589)
	(4342,3582)(4348,3582)(4348,3575)
\path(4058,3650)(4058,3644)(4058,3637)
	(4058,3630)(4058,3622)(4058,3616)
	(4058,3609)(4058,3603)(4058,3596)
	(4058,3589)(4058,3582)(4058,3575)
	(4058,3569)(4058,3562)(4058,3556)
	(4058,3549)(4058,3542)(4058,3535)
	(4058,3529)(4058,3522)(4058,3516)(4058,3509)
\path(4058,3509)(4065,3509)(4072,3509)
	(4072,3502)(4078,3502)(4078,3495)
	(4085,3495)(4085,3488)(4085,3481)
	(4091,3481)(4091,3474)(4098,3474)
	(4098,3468)(4098,3461)(4105,3461)
	(4105,3454)(4113,3454)(4113,3447)
	(4113,3441)(4120,3441)(4126,3441)
	(4126,3434)(4126,3427)(4133,3427)
	(4133,3421)(4139,3421)(4139,3414)
	(4139,3407)(4139,3400)(4139,3394)
	(4139,3387)(4139,3381)(4139,3374)
	(4139,3367)(4139,3360)(4139,3353)
	(4139,3347)(4139,3339)(4139,3333)
	(4139,3326)(4139,3319)(4139,3312)
	(4139,3306)(4139,3299)(4139,3293)
	(4139,3286)(4146,3286)(4146,3279)
	(4146,3272)(4153,3272)(4153,3265)
	(4160,3265)(4160,3259)(4166,3259)
	(4166,3252)(4173,3252)(4180,3252)
	(4180,3246)(4180,3239)(4186,3239)
	(4186,3232)(4193,3232)(4193,3225)
	(4193,3219)(4200,3219)(4200,3212)
	(4200,3206)(4207,3199)(4207,3191)
	(4207,3184)(4213,3184)(4213,3177)
	(4213,3171)(4220,3164)(4220,3158)
	(4220,3151)(4220,3144)(4220,3137)
	(4226,3137)(4226,3131)(4226,3124)
\path(4065,3677)(4058,3677)(4051,3677)
	(4045,3670)(4038,3670)(4032,3663)
	(4025,3663)(4025,3657)(4017,3657)
	(4011,3657)(4011,3650)(4004,3650)
	(3998,3650)(3998,3644)(3991,3644)
	(3991,3637)(3985,3637)(3985,3630)
	(3985,3622)(3977,3622)(3977,3616)
	(3977,3609)(3977,3603)(3970,3596)
	(3970,3589)(3970,3582)(3970,3575)
	(3970,3569)(3970,3562)(3970,3556)
	(3970,3549)(3970,3542)(3970,3535)
	(3970,3529)(3970,3522)(3970,3516)
	(3963,3516)(3963,3509)(3957,3502)
	(3957,3495)(3950,3488)(3950,3481)
	(3943,3474)(3943,3468)(3937,3461)
	(3937,3454)(3937,3447)(3929,3447)
	(3929,3441)(3923,3434)(3923,3427)
	(3923,3421)(3916,3421)(3916,3414)
	(3916,3407)(3916,3400)(3916,3394)
	(3910,3394)(3910,3387)(3910,3381)
	(3903,3381)(3903,3374)(3903,3367)
	(3903,3360)(3903,3353)(3897,3347)
	(3897,3339)(3889,3333)(3889,3326)
	(3889,3319)(3883,3312)(3876,3306)
	(3876,3299)(3869,3299)(3869,3293)
	(3863,3293)(3863,3286)(3856,3286)
	(3856,3279)(3856,3272)(3856,3265)
	(3849,3265)(3849,3259)(3849,3252)
\path(4065,3684)(4065,3677)(4058,3677)
	(4051,3677)(4045,3677)(4038,3670)
	(4032,3670)(4025,3670)(4017,3663)
	(4011,3663)(3998,3663)(3991,3663)
	(3985,3663)(3977,3657)(3970,3657)
	(3963,3650)(3957,3650)(3950,3644)
	(3943,3644)(3943,3637)(3937,3637)
	(3929,3637)(3923,3630)(3916,3622)
	(3910,3622)(3903,3616)(3897,3616)
	(3889,3609)(3883,3603)(3876,3603)
	(3869,3596)(3863,3589)(3856,3582)
	(3849,3575)(3841,3575)(3841,3569)
	(3835,3569)(3835,3562)(3828,3562)
	(3828,3556)(3822,3556)(3822,3549)
	(3815,3542)(3809,3542)(3809,3535)
	(3801,3535)(3801,3529)(3794,3529)
	(3794,3522)(3794,3516)(3788,3516)
	(3788,3509)(3788,3502)(3781,3502)
	(3781,3495)(3781,3488)(3781,3481)
	(3781,3474)(3775,3468)(3775,3461)
	(3775,3454)(3768,3454)(3768,3447)
	(3768,3441)(3761,3441)(3761,3434)
	(3754,3434)(3754,3427)(3748,3421)
	(3741,3414)(3741,3407)(3735,3400)
	(3735,3394)(3728,3394)(3728,3387)
	(3721,3381)(3721,3374)(3713,3374)
	(3713,3367)(3706,3360)(3706,3353)
	(3700,3353)(3700,3347)(3693,3347)
	(3693,3339)(3687,3339)(3687,3333)
	(3680,3333)(3680,3326)(3673,3326)
	(3673,3319)(3666,3312)(3666,3306)
	(3660,3306)(3660,3299)(3660,3293)
	(3653,3293)(3653,3286)(3653,3279)
	(3653,3272)(3653,3265)(3653,3259)
	(3653,3252)(3653,3246)(3653,3239)
	(3653,3232)(3653,3225)(3653,3219)
	(3653,3212)(3653,3206)(3653,3199)(3653,3191)
\path(4281,3663)(4274,3663)(4268,3663)
	(4261,3663)(4255,3663)(4248,3663)
	(4241,3663)(4233,3663)(4226,3663)
	(4220,3663)(4213,3663)
\path(4139,3684)(4146,3684)(4146,3691)
	(4153,3691)(4160,3691)(4173,3691)
	(4180,3691)(4186,3691)(4193,3691)
	(4200,3691)(4207,3691)(4213,3691)
	(4220,3691)(4226,3691)(4226,3684)
	(4233,3684)(4233,3677)(4241,3677)
	(4241,3670)(4248,3670)(4248,3663)
	(4255,3663)(4255,3657)(4261,3657)
	(4261,3650)(4268,3650)(4268,3644)
	(4274,3644)(4281,3637)(4288,3637)
	(4295,3637)(4301,3637)(4301,3630)
	(4308,3630)(4314,3630)(4321,3622)
	(4329,3622)(4335,3622)(4342,3622)
	(4342,3616)(4348,3616)(4355,3616)
	(4361,3616)(4369,3616)(4376,3616)
	(4376,3609)(4383,3609)(4389,3609)
	(4396,3603)(4403,3603)(4403,3596)
	(4409,3596)(4417,3596)(4417,3589)
	(4423,3589)(4423,3582)(4430,3582)
	(4430,3575)(4436,3575)(4443,3569)
	(4449,3569)(4457,3569)(4463,3569)
	(4470,3569)(4470,3562)(4477,3562)
	(4483,3562)(4483,3556)(4490,3556)
	(4497,3556)(4505,3556)(4511,3556)
	(4518,3556)(4524,3556)(4531,3556)
	(4537,3556)(4545,3556)(4552,3556)
	(4558,3556)(4565,3556)(4571,3556)
	(4578,3556)(4585,3556)(4592,3556)
	(4598,3556)(4605,3556)
\put(1411,3789){\ellipse{134}{134}}
\put(2516,909){\ellipse{122}{122}}
\path(281,3119)(2082,3129)
\path(1962.168,3098.334)(2082.000,3129.000)(1961.835,3158.333)
\path(311.000,4799.000)(281.000,4919.000)(251.000,4799.000)
\path(281,4919)(281,3119)
\path(291,3119)(1381,3789)
\path(1294.479,3700.603)(1381.000,3789.000)(1263.059,3751.718)
\path(1416.934,1907.984)(1387.000,2028.000)(1356.934,1908.016)
\path(1387,2028)(1386,199)
\path(1386,219)(3228,218)
\path(3107.984,188.065)(3228.000,218.000)(3108.016,248.065)
\path(1415,228)(2535,908)
\path(2447.995,820.079)(2535.000,908.000)(2416.856,871.366)
\path(3673.000,4814.000)(3643.000,4934.000)(3613.000,4814.000)
\path(3643,4934)(3643,3124)
\path(3641,3121)(5457,3129)
\path(5337.133,3098.472)(5457.000,3129.000)(5336.869,3158.471)
\put(61,4699){\makebox(0,0)[lb]{\smash{{{\SetFigFont{6}{7.2}{\rmdefault}{\mddefault}{\updefault}\normalsize{\it{y}}}}}}}
\put(3011,54){\makebox(0,0)[lb]{\smash{{{\SetFigFont{6}{7.2}{\rmdefault}{\mddefault}{\updefault}\normalsize{\it{x}}}}}}}
\put(1186,1859){\makebox(0,0)[lb]{\smash{{{\SetFigFont{6}{7.2}{\rmdefault}{\mddefault}{\updefault}\normalsize{\it{y}}}}}}}
\put(296,5174){\makebox(0,0)[lb]{\smash{{{\SetFigFont{6}{7.2}{\rmdefault}{\mddefault}{\updefault}\normalsize{$f(x)$}}}}}}
\put(1421,2264){\makebox(0,0)[lb]{\smash{{{\SetFigFont{6}{7.2}{\rmdefault}{\mddefault}{\updefault}\normalsize{$h(x)=f(x)*g(x)$}}}}}}
\put(1276,3444){\makebox(0,0)[lb]{\smash{{{\SetFigFont{6}{7.2}{\rmdefault}{\mddefault}{\updefault}\normalsize{$\delta(\vec{p})$}}}}}}
\put(2435,516){\makebox(0,0)[lb]{\smash{{{\SetFigFont{6}{7.2}{\rmdefault}{\mddefault}{\updefault}\normalsize{$\delta(\vec{p})$}}}}}}
\put(1133,140){\makebox(0,0)[lb]{\smash{{{\SetFigFont{6}{7.2}{\rmdefault}{\mddefault}{\updefault}C}}}}}
\put(0,3070){\makebox(0,0)[lb]{\smash{{{\SetFigFont{6}{7.2}{\rmdefault}{\mddefault}{\updefault}A}}}}}
\put(3382,3065){\makebox(0,0)[lb]{\smash{{{\SetFigFont{6}{7.2}{\rmdefault}{\mddefault}{\updefault}B}}}}}
\put(1867,2919){\makebox(0,0)[lb]{\smash{{{\SetFigFont{6}{7.2}{\rmdefault}{\mddefault}{\updefault}\normalsize{\it{x}}}}}}}
\put(5248,2919){\makebox(0,0)[lb]{\smash{{{\SetFigFont{6}{7.2}{\rmdefault}{\mddefault}{\updefault}\normalsize{\it{x}}}}}}}
\put(3453,4704){\makebox(0,0)[lb]{\smash{{{\SetFigFont{6}{7.2}{\rmdefault}{\mddefault}{\updefault}\normalsize{\it{y}}}}}}}
\put(3658,5179){\makebox(0,0)[lb]{\smash{{{\SetFigFont{6}{7.2}{\rmdefault}{\mddefault}{\updefault}\normalsize{$g(x)=\sum \delta$}}}}}}
\end{picture}
}

%% file: neuron_def.eepic
\setlength{\unitlength}{0.00043745in}
\begingroup\makeatletter\ifx\SetFigFont\undefined%
\gdef\SetFigFont#1#2#3#4#5{%
  \reset@font\fontsize{#1}{#2pt}%
  \fontfamily{#3}\fontseries{#4}\fontshape{#5}%
  \selectfont}%
\fi\endgroup%
{\renewcommand{\dashlinestretch}{30}
\begin{picture}(4377,3446)(0,-10)
\path(366,286)(1710,1630)
\path(1678.180,1492.114)(1710.000,1630.000)(1572.114,1598.180)
\path(1710,1630)(3059,1634)
\path(2939.089,1603.644)(3059.000,1634.000)(2938.912,1663.644)
\path(1720,1632)(2619,742)
\path(2512.615,805.105)(2619.000,742.000)(2554.828,847.745)
\path(1716,1630)(1715,2978)
\path(1745.089,2858.022)(1715.000,2978.000)(1685.089,2857.978)
\path(359,269)(359,3419)
\path(389.000,3299.000)(359.000,3419.000)(329.000,3299.000)
\path(360,281)(4365,281)
\path(4245.000,251.000)(4365.000,281.000)(4245.000,311.000)
\put(4278,54){\makebox(0,0)[lb]{\smash{{{\SetFigFont{6}{7.2}{\rmdefault}{\mddefault}{\updefault}\normalsize{\it{x}}}}}}}
\put(178,3139){\makebox(0,0)[lb]{\smash{{{\SetFigFont{6}{7.2}{\rmdefault}{\mddefault}{\updefault}\normalsize{\it{y}}}}}}}
\put(864,995){\makebox(0,0)[lb]{\smash{{{\SetFigFont{6}{7.2}{\rmdefault}{\mddefault}{\updefault}\normalsize{$\vec{s}$}}}}}}
\put(2283,1185){\makebox(0,0)[lb]{\smash{{{\SetFigFont{6}{7.2}{\rmdefault}{\mddefault}{\updefault}\normalsize{$\vec{d}_1$}}}}}}
\put(1823,2565){\makebox(0,0)[lb]{\smash{{{\SetFigFont{6}{7.2}{\rmdefault}{\mddefault}{\updefault}\normalsize{$\vec{d}_3$}}}}}}
\put(1814,3009){\makebox(0,0)[lb]{\smash{{{\SetFigFont{6}{7.2}{\rmdefault}{\mddefault}{\updefault}\normalsize{$D3=\delta( \vec{s}+\vec{d}_3) $}}}}}}
\put(3194,1599){\makebox(0,0)[lb]{\smash{{{\SetFigFont{6}{7.2}{\rmdefault}{\mddefault}{\updefault}\normalsize{$D2=\delta( \vec{s}+\vec{d}_2) $}}}}}}
\put(2734,649){\makebox(0,0)[lb]{\smash{{{\SetFigFont{6}{7.2}{\rmdefault}{\mddefault}{\updefault}\normalsize{$D1=\delta( \vec{s}+\vec{d}_1) $}}}}}}
\put(2683,1755){\makebox(0,0)[lb]{\smash{{{\SetFigFont{6}{7.2}{\rmdefault}{\mddefault}{\updefault}\normalsize{$\vec{d}_2$}}}}}}
\put(1407,1671){\makebox(0,0)[lb]{\smash{{{\SetFigFont{6}{7.2}{\rmdefault}{\mddefault}{\updefault}\normalsize{S}}}}}}
\put(0,299){\makebox(0,0)[lb]{\smash{{{\SetFigFont{6}{7.2}{\rmdefault}{\mddefault}{\updefault}\normalsize{A}}}}}}
\end{picture}
}

%% file: diagram.eepic
\setlength{\unitlength}{0.00061242in}
\begingroup\makeatletter\ifx\SetFigFont\undefined%
\gdef\SetFigFont#1#2#3#4#5{%
  \reset@font\fontsize{#1}{#2pt}%
  \fontfamily{#3}\fontseries{#4}\fontshape{#5}%
  \selectfont}%
\fi\endgroup%
{\renewcommand{\dashlinestretch}{30}
\begin{picture}(3968,4346)(0,-10)
\path(360,269)(1023,2735)
\path(1020.815,2611.326)(1023.000,2735.000)(962.872,2626.904)
\path(360,269)(360,4319)
\path(390.000,4199.000)(360.000,4319.000)(330.000,4199.000)
\path(360,269)(3956,263)
\path(3835.950,233.200)(3956.000,263.000)(3836.050,293.200)
\path(360,269)(2151,2959)
\path(2109.467,2842.488)(2151.000,2959.000)(2059.524,2875.740)
\dashline{60.000}(375,277)(803,2285)
\path(807.541,2155.202)(803.000,2285.000)(745.926,2168.335)
\path(1685.004,3280.093)(1713.000,3423.000)(1569.544,3397.965)
\path(1713,3423)(1029,2753)
\path(1708,3410)(1708,4083)
\path(1739.315,3957.750)(1708.000,4083.000)(1676.685,3957.750)
\path(1725,3420)(2398,3420)
\path(2272.750,3388.685)(2398.000,3420.000)(2272.750,3451.315)
\path(1711,3419)(2158,2966)
\path(2047.737,3033.159)(2158.000,2966.000)(2092.317,3077.149)
\path(2831.004,3492.093)(2859.000,3635.000)(2715.544,3609.965)
\path(2859,3635)(2175,2965)
\path(2853,3627)(3526,3627)
\path(3400.750,3595.685)(3526.000,3627.000)(3400.750,3658.315)
\path(2836,3623)(2836,4296)
\path(2867.315,4170.750)(2836.000,4296.000)(2804.685,4170.750)
\path(2851,3620)(3298,3167)
\path(3187.737,3234.159)(3298.000,3167.000)(3232.317,3278.149)
\dashline{60.000}(1484,2955)(1484,3628)
\path(1515.500,3502.000)(1484.000,3628.000)(1452.500,3502.000)
\dashline{60.000}(1485,2971)(2151,2971)
\path(2025.000,2939.500)(2151.000,2971.000)(2025.000,3002.500)
\dashline{60.000}(1486,2966)(1933,2513)
\path(1822.079,2580.563)(1933.000,2513.000)(1866.922,2624.812)
\path(1456.799,2824.753)(1483.000,2968.000)(1339.870,2941.168)
\dashline{60.000}(1483,2968)(803,2285)
\dashline{60.000}(2158,2279)(2156,2948)
\path(2187.877,2822.095)(2156.000,2948.000)(2124.877,2821.906)
\dashline{60.000}(2173,2292)(2839,2292)
\path(2713.000,2260.500)(2839.000,2292.000)(2713.000,2323.500)
\dashline{60.000}(2165,2292)(2612,1839)
\path(2501.079,1906.563)(2612.000,1839.000)(2545.922,1950.812)
\path(2135.657,2159.259)(2159.000,2303.000)(2016.432,2273.323)
\dashline{60.000}(2159,2303)(1496,1610)
\dashline{60.000}(362,272)(1496,1610)
\path(1441.299,1499.059)(1496.000,1610.000)(1395.527,1537.853)
\path(3265.428,1711.639)(3291.000,1855.000)(3147.989,1827.540)
\dashline{60.000}(3291,1855)(2613,1168)
\dashline{60.000}(3280,1849)(3278,2518)
\path(3309.877,2392.095)(3278.000,2518.000)(3246.877,2391.906)
\dashline{60.000}(3299,1838)(3746,1385)
\path(3635.079,1452.563)(3746.000,1385.000)(3679.922,1496.812)
\dashline{60.000}(3289,1849)(3955,1849)
\path(3829.000,1817.500)(3955.000,1849.000)(3829.000,1880.500)
\dashline{60.000}(356,261)(2613,1162)
\path(2512.675,1089.648)(2613.000,1162.000)(2490.430,1145.372)
\put(0,3914){\makebox(0,0)[lb]{\smash{{{\SetFigFont{9}{10.8}{\rmdefault}{\mddefault}{\updefault}\it{y}}}}}}
\put(2958,3673){\makebox(0,0)[lb]{\smash{{{\SetFigFont{9}{10.8}{\rmdefault}{\mddefault}{\updefault}$\vec{d}_{2}$}}}}}
\put(3172,3068){\makebox(0,0)[lb]{\smash{{{\SetFigFont{9}{10.8}{\rmdefault}{\mddefault}{\updefault}$\vec{d}_1$}}}}}
\put(2914,4175){\makebox(0,0)[lb]{\smash{{{\SetFigFont{9}{10.8}{\familydefault}{\mddefault}{\updefault}$\vec{d}_3$}}}}}
\put(2489,3413){\makebox(0,0)[lb]{\smash{{{\SetFigFont{9}{10.8}{\familydefault}{\mddefault}{\updefault}$\vec{s}$}}}}}
\put(3575,49){\makebox(0,0)[lb]{\smash{{{\SetFigFont{9}{10.8}{\rmdefault}{\mddefault}{\updefault}\it{x}}}}}}
\put(698,2607){\makebox(0,0)[lb]{\smash{{{\SetFigFont{9}{10.8}{\rmdefault}{\mddefault}{\updefault}$\vec{c}_1$}}}}}
\put(497,2069){\makebox(0,0)[lb]{\smash{{{\SetFigFont{9}{10.8}{\rmdefault}{\mddefault}{\updefault}$\vec{c}_2$}}}}}
\put(1244,1519){\makebox(0,0)[lb]{\smash{{{\SetFigFont{9}{10.8}{\rmdefault}{\mddefault}{\updefault}$\vec{c}_3$}}}}}
\put(2204,1088){\makebox(0,0)[lb]{\smash{{{\SetFigFont{9}{10.8}{\rmdefault}{\mddefault}{\updefault}$\vec{c}_4$}}}}}
\put(1512,2210){\makebox(0,0)[lb]{\smash{{{\SetFigFont{9}{10.8}{\rmdefault}{\mddefault}{\updefault}$\vec{p}$}}}}}
\end{picture}
}

%% file: circula.eepic
\setlength{\unitlength}{0.00030621in}
\begingroup\makeatletter\ifx\SetFigFont\undefined%
\gdef\SetFigFont#1#2#3#4#5{%
  \reset@font\fontsize{#1}{#2pt}%
  \fontfamily{#3}\fontseries{#4}\fontshape{#5}%
  \selectfont}%
\fi\endgroup%
{\renewcommand{\dashlinestretch}{30}
\begin{picture}(5759,6831)(0,-10)
\thicklines
\path(37,6446)(697,6446)(697,371)
	(37,371)(37,6446)
\path(37,1046)(727,1046)
\path(22,1721)(712,1721)
\path(22,2404)(712,2404)
\path(22,3079)(712,3079)
\path(22,3754)(712,3754)
\path(37,4414)(727,4414)
\path(30,5096)(720,5096)
\path(37,5771)(727,5771)
\path(1402,6431)(2062,6431)(2062,356)
	(1402,356)(1402,6431)
\path(1402,1031)(2092,1031)
\path(1387,1706)(2077,1706)
\path(1387,2389)(2077,2389)
\path(1387,3064)(2077,3064)
\path(1387,3739)(2077,3739)
\path(1402,4399)(2092,4399)
\path(1395,5081)(2085,5081)
\path(1402,5756)(2092,5756)
\path(2752,6416)(3412,6416)(3412,341)
	(2752,341)(2752,6416)
\path(2752,1016)(3442,1016)
\path(2737,1691)(3427,1691)
\path(2737,2374)(3427,2374)
\path(2737,3049)(3427,3049)
\path(2737,3724)(3427,3724)
\path(2752,4384)(3442,4384)
\path(2745,5066)(3435,5066)
\path(2752,5741)(3442,5741)
\path(5047,6431)(5707,6431)(5707,356)
	(5047,356)(5047,6431)
\path(5047,1031)(5737,1031)
\path(5032,1706)(5722,1706)
\path(5032,2389)(5722,2389)
\path(5032,3064)(5722,3064)
\path(5032,3739)(5722,3739)
\path(5047,4399)(5737,4399)
\path(5040,5081)(5730,5081)
\path(5047,5756)(5737,5756)
\thinlines
\put(2077.000,398.272){\arc{690.436}{0.0356}{3.1060}}
\path(2425.746,262.364)(2422.000,386.000)(2367.123,275.146)
\put(712.000,353.272){\arc{690.436}{0.0356}{3.1060}}
\path(1060.746,217.364)(1057.000,341.000)(1002.123,230.146)
\put(3427.000,383.272){\arc{690.436}{0.0356}{3.1060}}
\path(3775.746,247.364)(3772.000,371.000)(3717.123,260.146)
\put(1393.525,6482.478){\arc{647.174}{3.0286}{6.2568}}
\path(1662.109,6601.847)(1717.000,6491.000)(1720.730,6614.637)
\put(2714.500,6453.500){\arc{646.568}{3.0719}{6.2135}}
\path(2977.475,6584.429)(3037.000,6476.000)(3035.503,6599.684)
\put(5008.522,6452.522){\arc{647.178}{3.0285}{6.2570}}
\path(5277.110,6571.847)(5332.000,6461.000)(5335.731,6584.637)
\put(4657,3401){\ellipse{30}{30}}
\put(4657,3626){\ellipse{30}{30}}
\put(4657,3851){\ellipse{30}{30}}
\put(3757,3371){\ellipse{30}{30}}
\put(3742,3611){\ellipse{30}{30}}
\put(3757,3836){\ellipse{30}{30}}
\path(1057,3581)(1057,6461)
\path(1057,341)(1057,3566)
\path(1087.000,3446.000)(1057.000,3566.000)(1027.000,3446.000)
\path(2392,3566)(2392,6446)
\path(3757,326)(3757,3206)
\path(3787.000,3086.000)(3757.000,3206.000)(3727.000,3086.000)
\path(4657,4001)(4657,6431)
\path(2392,311)(2392,3536)
\path(2422.000,3416.000)(2392.000,3536.000)(2362.000,3416.000)
\end{picture}
}